# WEB 2.O For Small And Medium Sized Companies: A practical Case Study

Mark Abraham Magumba

**Abstract**:

*This project is a holistic analysis of web 2.0 and combines ideas from systems theory and organizational theory and goes on to attempt to bridge the gap between the ongoing theoretical discourse and its implications for the development process. The goal is to shift the discussion from a descriptive, highly philosophical one to a prescriptive framework that can directly be applied for the design, evaluation and deployment. Effectively this work posits that unless a practical framework can be synthesized from this rich debate the application and development of web 2.0 is at risk of proceeding haphazardly and this could potentially limit the benefits organizations reap from it. It then establishes the applicability of web 2.0 to the small organization and accordingly applies the framework to a practical case study.*

## Acknowledgements:

I would like to thank my family for all the support they have rendered to me throughout my studies

I would also like to thank Dr. Haemin Aziz for all the assistance he has rendered to me

Lastly I would like to thank Miss Terry Turner and Mr. David Gikawa for all the assistance theyôve rendered

## Table of Contents

*TABLE OF FIGURES*

*List of tables*

# Chapter 1 ï Introduction:

## 1.1. Overview

There has been a lot of research about the application of Information technology to business and this research suggests is a strong linkage between IT strategy and business success (Humphrey, 2001; Ward, Elm and Kushner, 2006; Mattison, 1999). However it should be noted that strategy in most cases is an evolutionary phenomenon and is the result of a continual process of trial and error, there are new challenges and requirements that emerge with time and what starts out as a single homogenous system morphs into a complex mix of technologies in a system of systems. Research is also abundant about the importance of obtaining an optimal fit between the strategy and the organization to which itôs applied ï the so called socio-technical fit (Stephen and Jian Cai, 1997; AppelBaum 1997). Though the logic is easy to follow, the practical implementation of these theories as noted by Appelbaum (1997) still poses a huge challenge, another thing as is evident from Cartelli (2007) who summarizes socio-technical theory as the view that the organization is essentially two subsystems, a social subsystem and a technical one that a systemôs view of the organization is necessary and it has to be acknowledged that the system exhibits synergies and emergent, often unpredictable behaviors.

Additionally the system like its external environment is primarily in a state of chaos (Galbraith, 2004; Hutton, 2003; McBride, 2005.) and quite inevitably a mismatch always exists between it and that environment. But by definition the organization exists solely to make profit (Williamson, 1964) therefore its most fundamental function is to sell which is hugely dependant on its interface or dialog with the customer. Thus ICT like other management tools acts as a regulating mechanism that attempts to minimize the chasm between the customer and organization. As regardless of its internal efficiencies, should the organization fail to provide an engaging interface between itself and the customer it will fail. Thus web 2.0 becomes an essential part of organizational ICT as if correctly applied it can enrich that interface and reduce the gap between the organization and its customers thereby increasing the customerôs long term value.

This research is about the applicability and eventually the application of web 2.0 to SMEs. The sponsoring organization Key Maid cleaning organization, a sole proprietorship also doubles as the case study.

## 1.2. Defining SMEs:

According to Rowe (2008) Small and medium sized enterprises (SMEs) are the backbone of the UK economy through their significant contribution to its productivity and performance. Lawless, Allan and OôDwyer (2000) described SMEs as comprising micro businesses with less than 10 members of staff, small businesses with 11 to 50 members of staff and medium businesses with 51-250 members of staff, this is also the European Commission Definition. Choueke and Armstrong (2000) posit that there exists some difficulty in defining what a business is and as a consequence Choeke (1992) defined a small and medium sized enterprise as any enterprise that categorizes itself so. Storey (1994, page 8) rightly acknowledges that there is no standardized definition for a small firm. Though size is a common feature in most definitions of small firm (Storey 1994, Bolton Committee Report 1971, European Commission Definition also the Department of Trade and Industry definition), Wilkinson (1999) observes that there is also little consensus on the definition of small. According to the Department for Business, Enterprise & Regulatory Reform (BERR)ôs Enterprise Directorate Analytical Unit, in 2007, the UK economy was 99% SMEs (Rowe, 2008). Lawless, Allan and OôDwyer (2000) also reported that 99% of organizations in the EU are SMEs majority of which are micro organizations with less than or up to 10 employees and further added that these contribute 66% of the overall GDP. The importance of these small firms therefore cannot be underrated. Taylor and Murphy (2004) put the GDP contribution at 40% and further claim SMEs employ 12 million, have an annual turnover of approximately 1 trillion pounds and employ about 55% of the private work sector.

## 1.3. *The Research:*

The research aims to establish the applicability of web 2.0 concepts to Small and Medium sized businesses through a broad theoretical review. From this theoretical investigation I hope to draw conclusions that can be applied to practical case and derive an implementable design. The specific objectives are outlined below:

i. Review the development of web 2.0 and evaluate the current industrial view points on this family of technologies

ii. Demonstrate the criticality of web 2.0 technologies through a thorough theoretical review of organizational theory and current technological trends

iii. Demonstrate the relevance of web 2.0 to small and medium enterprises by identifying opportunities for implementation of web 2.0

iv. Apply the theoretical conclusions to a practical business case to evolve a model for implementation of web 2.0

So, ultimately the goal is to map theory to practice in somewhat explicit terms to overcome the vagueness and generalizations that are typically characteristic of investigations into web 2.0.

## 1.4. *Structure of the Thesis:*

Chapter 2 describes the background story of SMEs and the initial version of the world wide web and deals with web 1.0 shortcomings by application of a systems theory perspective, chapter 3 is dedicated to modeling the web 2.0 problem and implicitly presents it as the product of the natural evolution of efforts to manage the classical management problem, characterized as a continuous, self adjusting, iterative quality maximizing and cost minimization loop; chapter 4 employs organizational theory to argue the case for employing web 2.0 to the small and medium sized enterprise, chapter 5 tackles web 2.0 in the wider context and sets the stage for the next chapter by deriving global conclusions for the discussion to that point, chapter 6 uses those conclusions to derive a prescriptive outcome for the practical application of web 2.0 to SMEs, chapter 7 is an evaluation of the overall project and chapter 8 presents the overall conclusion and recommendations. Appendices 1 to 6 are the deliverable to the client. Appendices 1 to 6 outline the deliverable to the client. Appendix 1 outlines the background, business opportunity, and customer Needs, Appendix 2 the solution features, appendix 3 the business context, appendix 4 the business rules, appendix 5 the Systems Requirements Specification and appendix 6 is the budget plan.

# Chapter 2 ï The World Wide Web and SMES

### 1.5. 2.1. Introduction

Many SMEs fail and the most frequently advanced reasons are in the neighborhood of managerial deficiencies (Kent 1994, Duchesneau and Gartner 1990 p.309). Managerial talent could originate from within the organization or be the result of external consultations and in both cases should the managerial mechanism be deficient, the business is likely to fail. The obvious problem this presents for firms is how are they supposed to identify good advice and even if they do are they able to afford this good advice? Not many firms categorized as small for example are capable of maintaining specialist management roles within their portfolio to cater to marketing, financial and planning functions yet these are required by any organization to survive. Therefore the business owners find themselves in multiple roles as company accountants, advertising executives, and financial managers and many a time they do not have the specialist knowledge required for these roles rather they are dependent on their natural instinct. Watson (2003) determines that businesses that seek external advice particularly from others within the business and accountants particularly had low failure rates. Whereas it could be argued that even the accountants themselves or industry experts do not necessarily employ discrete rules and are also largely dependent on their instincts, figuratively speaking ninety percent of the time youĝll find that experts have a tendency to follow rules as naturally individuals are risk averse. The utility of ICT therefore has to be demonstrated in its ability to codify these rules (cum business rules) and deliver them through a simple interface to business owners short of that ICT can re-define the communication infrastructure required to access this information by reducing the cost of information and shortening the communication pathways.

### 1.6. 2.2. The Digital SME

Taylor and Murphy(2004) note that SMEs are progressively engaging with the Knowledge economy through connection to the internet, use of internet brochures, construction of e-commerce enabled websites and the full integration of their websites with their back office operations. They also assert the view that the adoption of ICT is inexorably linked to competitive pressures thus ICT is a route to greater efficiencies. An article by bytestart.com notes that UK small business owners work approximately half a billion man-hours a year which in comparison to an employee working 35 hours a year represents two extra hours a day which demonstrates that like any other business owner small business owners are committed to success. However, this commitment is not always coupled with the appreciation of the power of ICT. This in addition to the fact that ICT is normally associated with huge investment and these have to go hand in hand with creative, informed staff in order to fully exploit and to begin with the firmĝs employees have to be computer literate may slow the adoption of ICT.

The World Wide Web promises low Online Management System business solutions for SMEs. Web 2.0 refers to a group of technologies aimed at enhancing the functionality of the web. As ambiguous as this sounds it goes as far as to point out that web 2.0 is a trend rather than a specific technological achievement. Central to this theme is the increasing intelligence of web applications as well as their portability. It is more representative of the way in which computing is applied as opposed to a technological breakthrough as many of the components that are the foundation of web 2.0 have actually been in place since the inception of the world wide web. So, as part of that change is an enhanced understanding of the requirements for success in a connected environment. Bruno-Britz (2008) contends that web 2.0 is less about technology than it is about creating utility for users. This is in close agreement to previous computing literature that for a long time has almost unanimously viewed requirements analysis as the most important part of applications design, a view that resonates with the contemporary view of marketing. Although sadly in the traditional design approach web applications were built with little internal flexibility to enable the user to meaningfully impact the vital aspects of quality.

## *1.7. Systems Theory Perspectives:*

As earlier acknowledged the firm is an open system that is primarily a state of chaos meaning that its state is dynamically changing and it never achieves a true stability but rather tends to points of relative stability also known as strange attractors. The management function operates intuitively to stop the organization from decaying into a total state of chaos in order to produce predictable outcomes within a changing environment. But even more problematic is the dynamic nature of the external environment which dictates the organizationâs success criteria. The external environment presents management with the following system dynamics challenges:

### 1.7.1. Partial Control of Events and trends

The managerial function cannot directly control it and it can shift and change independent of internal organizational efforts. For example the failure of a new product line may be entirely independent of internal strategic decisions and simply be down to the arrival of a new competitor. Thus marketing through its ability to influence customer behavior becomes invaluable to all businesses (Kjaer-Hansen, 1967). The important thing therefore is for a business to have the right level of skills and internal flexibility to immediately reconfigure itself and change the important aspects of its marketing mix (DâAveni, 1998). As a result of this marketing has come to exert more and more influence on cooperate strategy and represent a bigger fraction of cooperate expenditure (McIntyre, 1990) as it could make the difference between failure and success. Increasing the efficiency and effectiveness of this process has long been the emphasis of the internet and one of the driving factors behind its success. The internet (web 1.0) solved some of the oldest marketing dilemmas by being able to transcend barriers of age, religion and language through its wide reach and its programmability but not till the advent of web 2.0 has it even come close to solving the fundamental marketing dilemma by enabling the customer to directly express their opinion on quality and to a certain extent dictate the quality attributes of products or services. The ability of clients to co-author content, rate services and review products via wikis, blogs and content sharing sites like youtube has influenced the way in which services are delivered and many businesses are moving more towards this democratic model of marketing. A good example is Dell corporation which enables a buyer to configure their computer online, their website then calculates the price of the unit based on the chosen components and the order is passed on to the production line and a computer fitting the buyerôs exact specifications is shipped to the buyer. This is a fundamental shift in marketing practice as it is no longer a case of manufacturers trying to guess what the markets want but rather a case of businesses asking the markets on a per customer basis what they want and giving each customer exactly that.

### 1.7.2. A Dynamic Problem and Solution Space

The speed of change is another debacle for the organization. Like the internal system the external environment comprises complex interactions and exhibits emergent behaviors albeit on a grander scale but unlike the internal organizational space it is ungoverned and unstructured. This coupled with the fact that organizational systems may be fairly unstable, especially for organizations in the initiation or growth phase of its life cycle, that is a small change in conditions can result in a large change in the behavior of the system (Mesarovic and Takahara, 1975). This instability is even worse in the external environment and the effect of changes is highly unpredictable. So, managers find themselves suspended between two systems that are largely unquantifiable that is they canôt be expressed as a finite set of variables connected via discrete relationships meaning in the absence of business intelligence management is a merely a blatant guessing game. Another issue that emerges here is the fact that even for organizations with the capabilities to gather this business information it has been extensively documented that the dissemination of that knowledge through parts of the organization could be difficult meaning that there is still a huge likelihood that despite having access to critical information organizations may remain unresponsive (Rynes et al, 2001). There exists documentary evidence linking better knowledge transfer to competitive advantage (Szuslanski, 1996; Dayasindu 2002) and some observers have come to point out that a lot of the information that has been accumulated by management science is impractical (Bayer and Trice, 1962, pg. 608). So, effectively in the traditional format there are two levels of distortion. Thereôs information distortion between the research community and the user community and additional distortion occurs between the research community and the organizational practitioners. Besides the possible distortions that may occur within this chain of information transfer, thereôs the issue of delay. The organization would need to have a very high level of agility in order to apply new knowledge quickly enough and avoid losing competitive advantage as would result from a mismatch between organizational strategy and the organizationôs environment (Venkantraman and Prescot, 1990).

### 1.7.3. Quality Assurance in a Dynamic and Competitive Market Space

As a consequence of this continual and rapid change, thereôs also a changing definition of quality so at any point in time a cognitive gap exists between the organizationôs view of quality and the external environmentôs view. At this point it is important to attempt a definition of quality. Russell and Miles (1998) posit that quality can be attacked both perceptually and conceptually. The perceptual elements of quality imply that to a certain extent quality is an individual concept and relies upon personal opinion. So, effectively in the pursuit of quality the organization is in a continuous state of negotiation with the customer and the external environment. Berry and Waldfogel express this mathematically as follows:

$u_{ij}$ *being the utility of product j for the* $i^{th}$ *customer*

$$u_{ij} = I_{ij} Ö_j - P_j \ldots \quad (1)$$

Where P is price, assuming away income effects, and measuring utility in pounds, **I** is the consumerôs willingness-to-pay for quality and is distributed on the interval (0, ∞) so that there are some consumers with arbitrary high value of **I** who will pay for an increase in quality ($Ö_j$) to any level. Berry and Waldfogel effectively posit that utility is negatively influenced by price but also importantly that where **I** tends to infinity the role of price can become peripheral. That is a commodity can be so appealing to a buyer to the extent that they are willing to pay an exceedingly high price for it. Also of great importance this model accounts for competition by using the change in quality rather than quality itself therefore can be used to explain the influence of making changes to quality on market success in a competitive space. Furthermore, Shaken and Sutton (1987) aptly observe that quality improvements may be brought about by increasing variable costs or by fixed costs. Combining these two ideas it would seem that in order to deliver high quality to consumers the business risks lowering the utility of the product by increasing the price. As presented in this formula it is apparent that quality and utility are personal concepts and therefore vary from individual to individual which automatically invalidates traditional market wide definitions and advocates for more intimate, personalized campaigns via customization as noted by Tseng and Piller (2003).

The organization seeks a compromise between the customerôs required quality parameters and its internal capabilities like technical ability, supply chain efficiencies and internal revenues. So, the ability of technology to increase efficiency and tame costs cannot be underestimated in a competitive market where businesses are continually aiming to differentiate themselves on quality. The need for differentiation requires a constant dialog with the customer and web 2.0 approaches ensure that that dialog is a genuine two way conversation; Piller, Stotko and Moeslen (2004) refer to this dialog with the term ñcustomer integrationò, it also requires the organization to undergo a paradigm shift by employing information more intuitively to target customers more intelligently at a fraction of the cost.

### 1.7.4. The Customization Challenge:

The lack of structure of the external environment within which the organization operates means itôs impossible to evolve an accurate or even meaningful generic model and it is better interpreted not as a single system (mass production) but rather as several microscopic systems (individual stakeholders) each engaging with the organization at their own unique interfaces (mass customization). Therefore for the competitive organization, change cannot be interpreted generically via aggregated functions but rather a greater specificity is required to capture the full complexity of these unique, autonomous systems. This is only achievable via enhanced data collection and analysis techniques thus data, particularly micro data (data on an individual) as opposed to macro data (aggregated data), is becoming more and more central to operating on the web. However, of even more vital importance is the ability to transform data into actual information. OôReilly (2005) argues quite explicitly that a given software product is only as important as the amount and variability of the data it helps to manage. In other words, if the web were an organism, then data is the cell structure of the organism and information is the sum total of its physiological function. This positive relationship between information, data and utility means quite simply that at a fundamental level organizations must maximize the amount of data in their databases. Web 2.0 provides mechanisms not only for data maximization but also maximization of utility by maximizing information through remixable data sources and the fuller integration of intelligent techniques. On the other hand this data explosion imposes a steep hardware requirement on the organization for example Anderson (2006) reports that Googleôs database is now in the order of hundreds of penta bytes ($10^{15}$ bytes) thus for the small business this sort of investment in hardware and technical competencies is simply infeasible and this inevitably will drive the rise of web services.

### 1.7.5. Intimacy versus Aggregation:

The true nature and internal composition of these autonomous, microscopic systems (individual stakeholders) is unknown and is simply inferred from their behavior. The internal cognitive apparatus of individuals is a mystery that defies analysis though it is widely thought to be a rational profit maximization function, that is individualôs pursue the course of action that promises maximum payoff. However, Simon (1978, page 5) notes that thereôs little positive evidence supporting this theory and argues that most of the evidence actually points to bounded rationality rather than utility maximization. As a consequence of this fact managerial efforts are essentially reactionary. Even when management apparently anticipates the markets for instance, in actuality it most certainly will be reacting to documented signals derived from data from past observations. However, when your view of the market is an aggregated sales figure it takes a while for the numbers to accumulate so thereôs always the likelihood of reacting too late to an ongoing trend. Internet technology on the other hand gives business the ability to respond to minute events like mouse clicks by a singular customer instance in real time.

However, despite all of this aggregation is important as individual choices are also influenced by trends thus business requires a big picture perspective. It has been experimentally demonstrated that indeed subtle cues in a customerôs environment can influence purchase decisions (Dijksterhuis et al, 2005 page 2). Thereôs a wide range of variables and the relationship between micro and macro events may not be obvious yet more often than not it is significant. So, the great organizational dilemma is how to operate on those aggregate figures whilst maintaining an intimate relationship with individuals. Rockwell Automation (2008) puts forward the idea of ñTotal Valueò which is a value definition arrived upon by taking into account fiscal, utility and importance dimensions. This undoubtedly requires a total view which can only be synthesized by aggregating input from several stakeholders. Additionally Rockwell call for a thorough understanding of business processes which is especially important in consideration of the fact that as the organization grows it becomes more of a system of systems (Lisa et al, 2006) as management becomes more and more involved with maximization of synergies between these various systems.

### 1.7.6. Interoperability and Systems Integration (I&I) Challenges:

Lisa et al (2006) go on to posit that ultimately traditional methods are inadequate in the management of such systems and propose a novel technique to address the interoperability challenges posed by such a system. Also importantly they note that in the systems of systems (SoS) parlance interoperability goes beyond the ability of technological components to seamlessly work together but also the integration of procedures, human resource and even organizational politics for example Morris et al (2004) note that a lot of the barriers to system integration are actually programmatic. This wider definition of interoperability is especially relevant because now more than ever integration is important as naturally systems are developed and tested in isolation but they must interact with a multitude of other systems and at times they interface with these systems in ways that are less obvious so integration is becoming an increasingly significant factor in the success of information systems and a source of competitive advantage for the organizations as a whole. There are also fiscal arguments for greater systems integration for example according to Brunnermeier and Martin (1999) poor data integration cost the US automotive industry a billion dollars a year in lost revenue.

If the small organization is interpreted as an organization in its early inception, then it should be assumed that should it survive it will inevitably grow in size and revenue therefore in anticipation of this it is important to pay particular attention to the organizational architecture and processes as well as the technology. The shift from web 1.0 to web 2.0 is especially the result of Information Systems architects paying a greater attention to streamlining of business processes and achieving a tighter fit between strategic objectives and system capabilities.

# Chapter3 ï Modeling the Web 2.0 Problem

### *3.1. A Mathematical Model Depicting the Management Process*

By definition the organization exists for some purpose which reveals itself in the set of activities that its human component undertakes. If say an Information System view of the organization is employed, this reality is captured explicitly in the definition which posits that the Information System is a combination of people, processes and technology. The human component organizes the technology component and itself so as to achieve the organizationõs function by adjusting the characteristics of that process. Given the organization is financial firm that exists primarily for profit then at any instant in time there is an optimal organizational configuration of resources for the maximization of profits. This optimum configuration is variable with time as the organization operates within a dynamic environment.

If say ***f (t)*** is the function that links the organizationõs resource mix at a time t, to a particular quality attribute for a given product and ***g (t)*** is the actual function that reflects the internal resource mix that returns the desired quality attributes for a given product for a given customer over time. The gap between the organizationõs value proposition and the external environmentõs quality interpretation is represented by the equation

$$Q_g = g(t) - f(t) \ldots\ldots\ldots\ldots\ldots\ldots\ldots (i)$$

Following this logic the management function can be expressed as

$$Minimize\ (g(t) - f(t)) \ldots\ldots\ldots\ldots\ldots\ldots. (ii)$$

In other words management is the attempt to minimize the gap between the customerõs desired quality interpretation and the organizationõs value proposition. The management function operates as a feedback loop and is controlled by indirect data with a known correlation with quality like actual sales figures. A difference between the actual sales and the sales targets, or rather a non-zero value of $Q_{g,}$ indicates a mismatch between the internal interpretation of quality and the interpretation of the target market. If $Q_g$ is positive that is if the clientõs expectations are higher than the internal organizational view, then actual sales lag target sales and the inverse too is true. Ideally the organization aims for negative values of $Q_g$ equivalent to the organization surpassing its targets. Much of the organizations product development will be concerned with determining which attribute has the strongest influence on the productõs quality or more importantly **g (t).** Figure 1 below graphically illustrates this idea and figure 2 applies real data to further demonstrate it

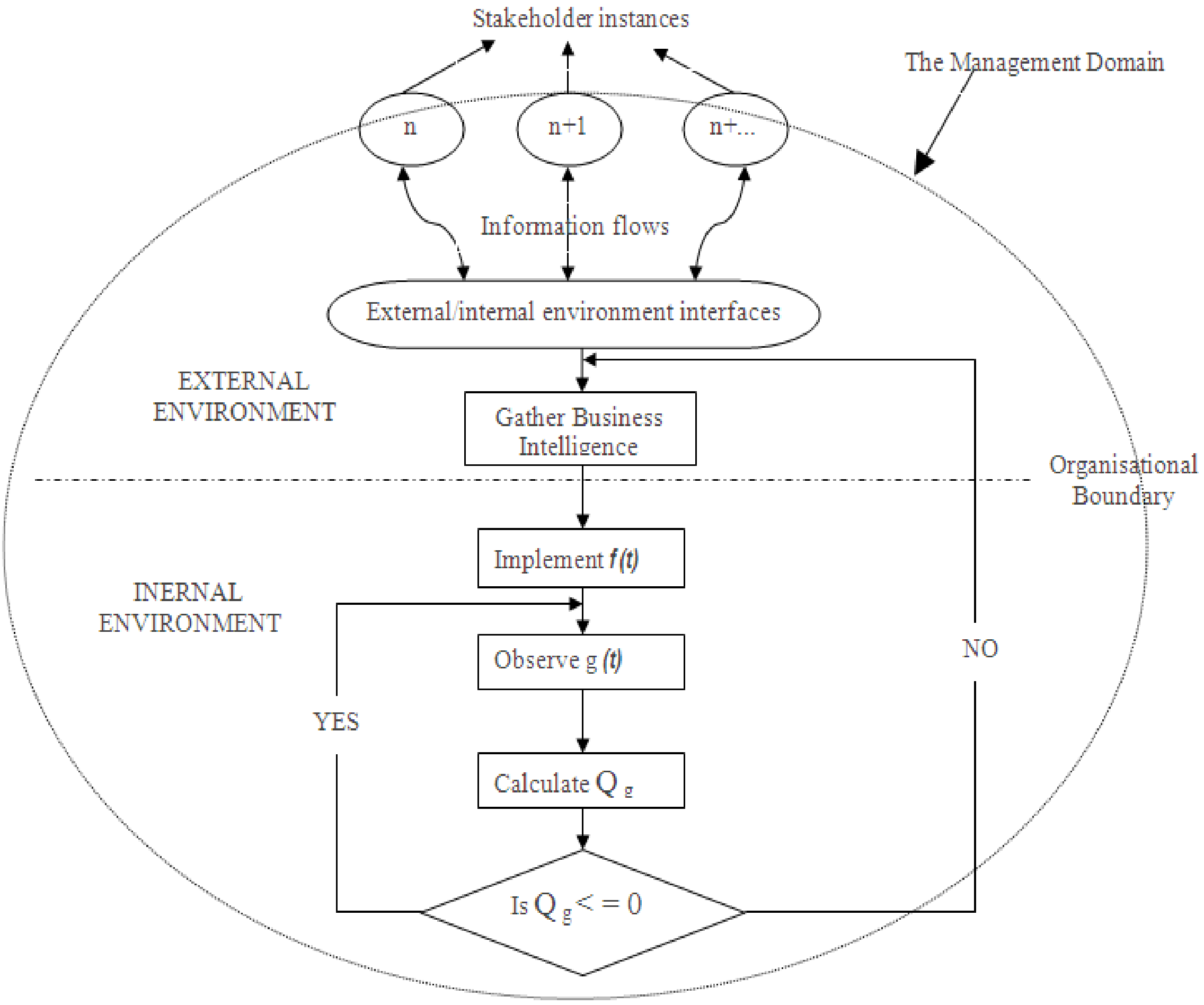

*Figure 1 Flow Diagram Depicting the Management Process as a continuous and iterative minimization of $Q_g$*

- 2004: 509mt
- 2005: 630mt
- 2006: 750mt

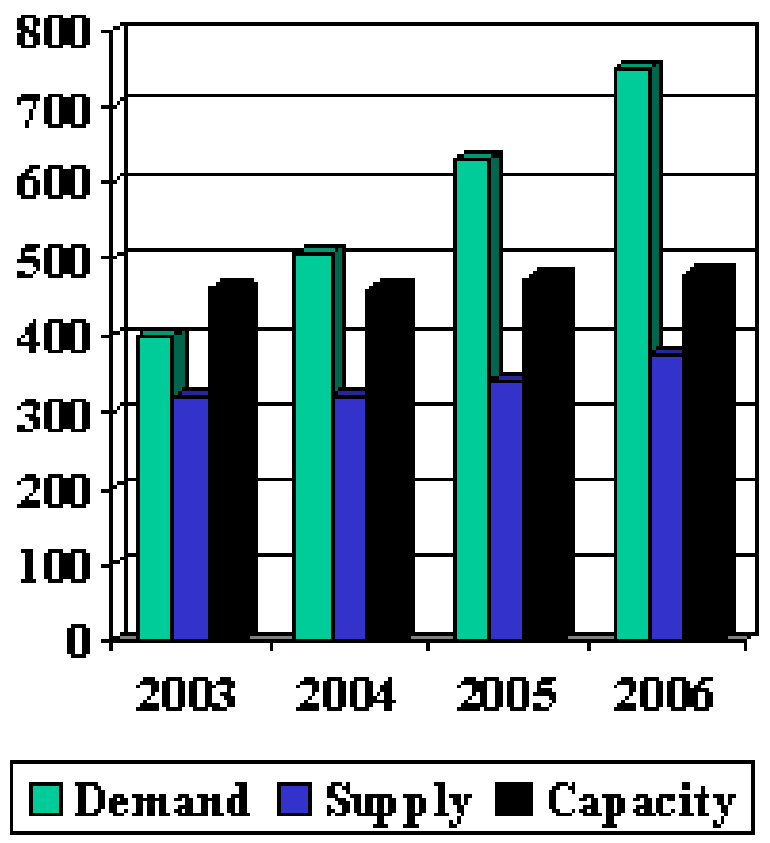

*Figure 2: Bar graph depicting global indium demand against supply against capacity from 2003 to 2006 ( O'Neil (2004))*

Figure 2 above demonstrates how the managerial loop operates employing data from global indium[1] production. In 2003 the supply lagged both demand and production capacity which can be interpreted as an anecdotal evidence of the fact that there was a mismatch between f(t), the supplying organizationsô interpretation of the global value hence demand for indium and g(t), the actual global value and hence demand for indium. Even though it is evident throughout the data that supply lags demand and production capacity but starting 2005 a slight adjustment in supply is evident which can be interpreted as the suppliersô managerial mechanism adjusting f (t) to minimize the term g(t) ï f(t) or the quality mismatch. It should be noted that other factors might affect the managerial response for example as seen here the supply is consistently below the actual capacity even though it lags demand which could be the suppliersô way to exacerbate the deficit and artificially keep prices high.

It is particularly useful to interpret the managerial process as a looping function constantly trying to minimize a random $Q_g$. As a loop it has certain properties which include:

### 3.1.1. Frequency of the loop

This would refer to the number of iterations the loop undergoes within a given period of time. Traditionally this would normally correspond to the period of time between successive management reviews however typically the rules change in drastic periods for example when addressing emergencies. A fair amount of energy and managerial resources are expended in preparing reviews and acting on them therefore the higher the frequency of the loop the higher the managerial overhead and the higher the resource requirements. It is important to note that typically there are several loops co-existing within the overall managerial process corresponding to the different levels of management for example one loop could be an accounting period and another loop could be an inventory replenishment period.

### 3.1.2. Sensitivity of the loop

The period of the loop neednôt be fixed and in most practical scenarios it is not. Most management activities are in fact triggered by situations. Sensitivity here refers to the extent to which a given situation or set of variables has to deviate from the normal or desired configuration for management to initiate corrective action. A higher sensitivity means that thereôs a small delay between the onset of change and corrective action therefore this minimizes the impact of such changes. It is also clear here that the smaller the units of measurement the higher the sensitivity so the organization must always aim to operate in the smallest possible units for example the organization should have the capacity to react to trends at the level of an individual client or employee.

[1] Indium is an element used in the production of Liquid Crystal Displays (LCDs)

### 3.1.3. Accuracy of the loop

This refers to the error of adjustment and is the difference between the calculated $Q_g$ and the actual $Q_g$. This like in any other loop is closely related to the sensitivity of the loop and its frequency. That is the higher the sensitivity the higher the accuracy but the quality of the algorithm used to calculate $Q_g$ also strongly influences the accuracy of the outcome and needless to say this is a changing algorithm as management practice is dictated by a dynamic and highly stochastic environmental situation.

### 3.1.4. Period of the Loop

This refers to how long an iteration of the loop would take. It is especially important in error conditions, that is, when the loop is triggered in response to an error that requires correction. The longer the loop takes the longer the error condition persists and the higher the cost to business. Take for instance when a company fails to meet demand for a given product. The longer it takes the company to restock and cater for waiting customers the more damage it does to its reputation and beyond a certain period it can potentially suffer lasting damage as a result.

Crucially, the model shows the management domain overlapping into the external environment. Management is not merely a reactionary sport but through advertising and marketing engages the external environment and tries to influence stakeholderȏs opinion. The importance of this model is that it synthesizes a relationship between managerial practice and quality and therefore establishes a transitive relationship between organizational processes and stakeholders.

# Chapter 4 ï The Web 2.0 Movement

## *4.1. Overview*

So far the argument has been made for web 2.0 but an in depth discussion of web 2.0 hasnòt yet been ventured into therefore this section shall be devoted to providing a review of web 2.0 as a technology and as a concept. The term first used by Darcy Dinucci (Dinucci, 1999) to describe the ongoing evolution of the web from a static, unidirectional model to a more interactive space. Oberhelm (2007) defines web 2.0 as a set of design principles and tools that are designed to enable users to ñcomment, collaborate and edit informationò rather than ñserveò information from authorities to a receptive, passive audience. This is the predominant view and is shared by several commentators such as Tredinnick (2006), Harrison and Brethel (2009), Thackery et al (2008) and Hardey (2008), however this view is not without its detractors. The use of the term web 2.0 has also been challenged by some commentators for example Berlind (2006) challenges the underlying logic behind the use of the term citing AJAX (Asynchronous Javascript and XML) as the only significant technological change that has occurred since the inception of HTTP (Hyper Text Transfer Protocol). Others like Laningham (2006) have argued that web 2.0 is an unnecessary terminology because in actuality it is in fact a realization of the original vision of Tim Barnes Lee the originator of the World Wide Web (web 1.0). Proponents for the term also agree but they attribute the persistence of the term more and more to the way internet technology is applied than to the technology itself. OòReilly (2005) to whom the terminology is widely attributed points out seven core competencies which include:

### 4.1.1. The use of the internet as a platform:

This refers to the seamless integration of two or more websites to create a unique service. A good example is facebook.com. Facebook.com released details of its Application Programming Interface (API) enabling independent programmers to produce applications for facebook. Users on facebook can seamlessly download and ñinstallò applications to their profile but unbeknownst to them all these applications are in fact hosted on different servers that have been enabled to communicate with the facebook servers via a series of parameters like application Ids and specialized php or JavaScript libraries. Many companies are beginning to embrace this model of operation and the most notable include Google and yahoo.com

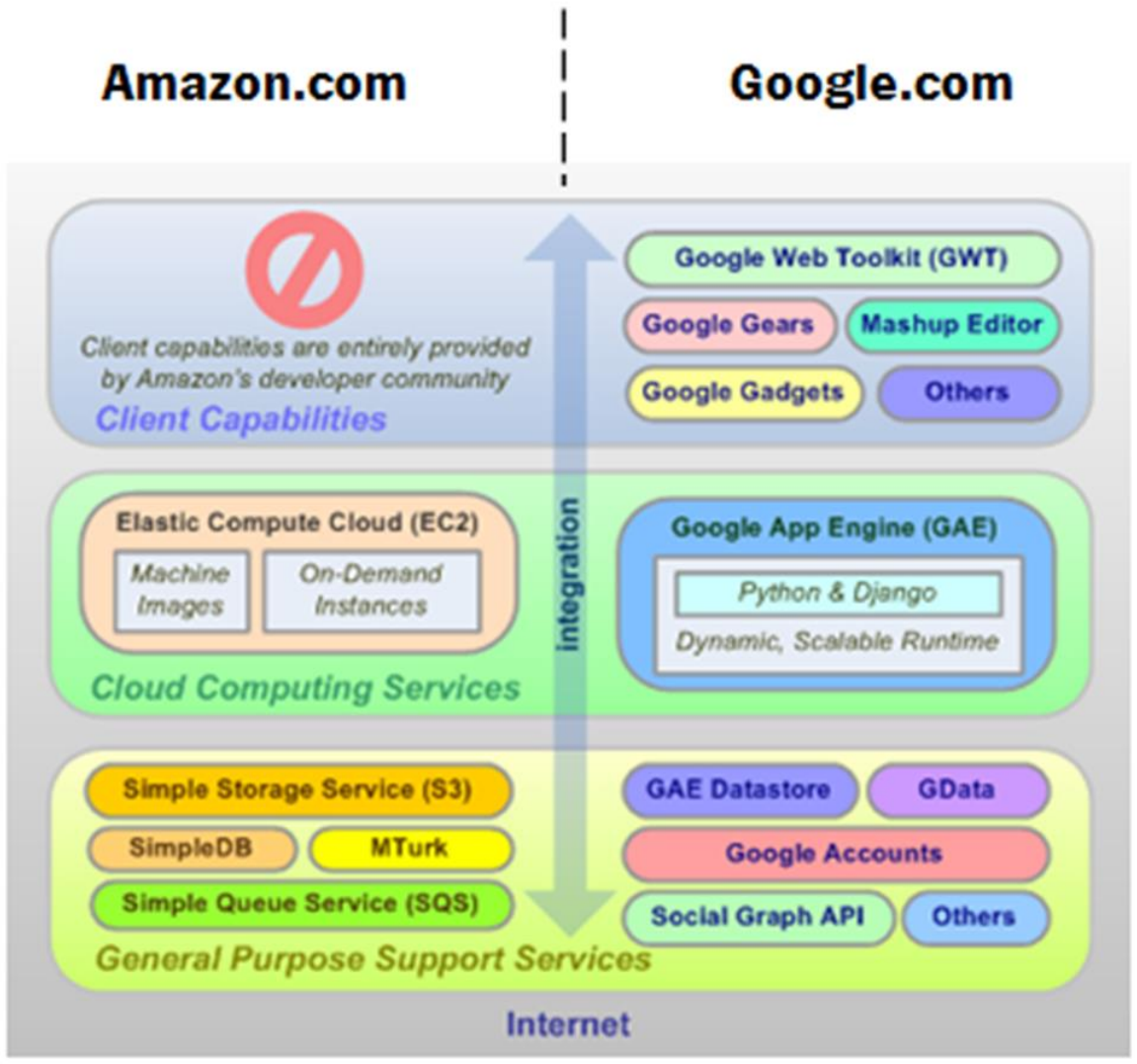

*Figure 3: Comparison between Amazon.com and Google Platforms: Source: Hinchcliffe (2008)*

Figure 3 above demonstrates this idea by comparing two leading internet platforms namely amazon.com and google.com. In both cases the systems are architected into three interrelated tiers namely the client capabilities, cloud computing services and general purpose support services. This is possible because the system is architecturally implemented as a MVC (Model-View-Control pattern), an architectural approach that facilitates multiple ways to view the same data by separating data (models) and the means to display it (views) and the means to affect the data and its display (Control). Figure 4 below depicts the MVC abstraction

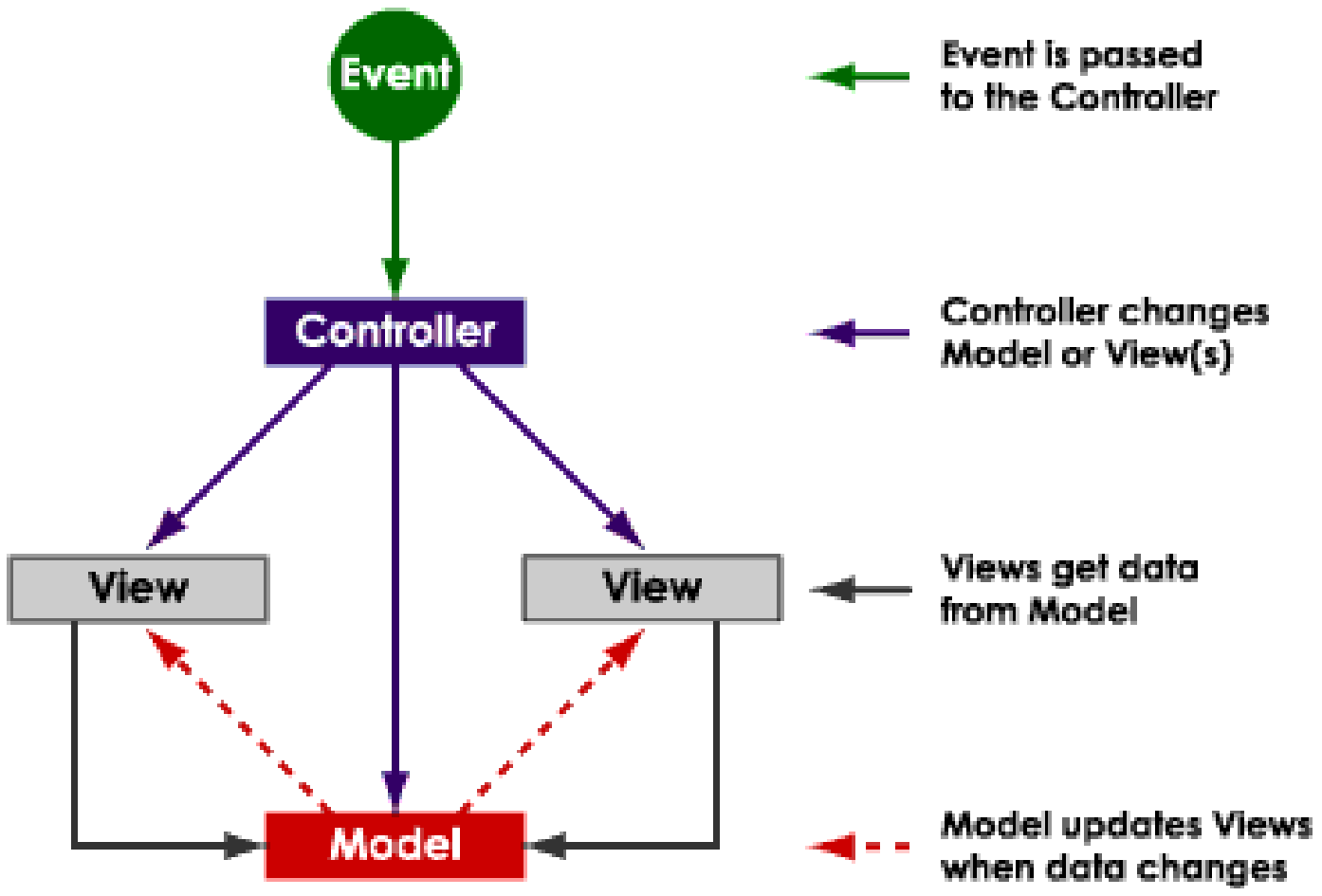

*Figure 4: Model-View-Control Applications Framework Platform Abstraction,*

*Source eNode (2002)*

The implications of this is that several services can be implemented using the same data by manipulating it differently by the controllers. The API operates at the control level of the framework.

### 4.1.2. User generated Content

This refers to the decentralization of the power to author content. The most explicit adoption of this concept is embodied by wikis which are sites that offer users the power to edit the appearance and content of web pages. Although a majority of social networking websites like facebook, youtube, myspace will enable users to upload content, provide mini blogs, podcasts and other features that enable users to author content.

### 4.1.3. Re-mixable data source and data transformations (“mash ups”)

This is the idea that services are architected as portals aggregating and recycling data from an array of disparate sources and concentrating it at one point for the user. There are special utilities known as aggregators. A very common type of aggregator is a feed aggregator which combines RSS and ATOM feeds from several services into a single service.

### 4.1.4. Software portability over multiple devices

This is a reference to the property of a service to run seamlessly across multiple devices. A good example here is micro formats which are open data formats. They are typically implemented in DHTML with special div tags and they effectively slice up the content using metadata descriptors. Examples include hCard (for contact information), hCalendar (for events) and hAudio (for audio files). By way of a specialized plugin like operator or some programming library like JQuery (oomph reader) micro formats can be read by the client computer and this data extracted. For example the micro format reader can extract all contact information and sync it to a phone, or it can extract all multimedia and Bluetooth it to a compatible device.

### 4.1.5. Harnessing the power of the crowd. (“Collective Intelligence”)

This is the reliance on the user communityôs inherent cognitive capability to define and create value. This is achieved by providing the users with the capability to author content as well as rate it ï it is then generally assumed that popular content is of more value to the community of users. Another example is the concept of ñfolksonomyò where tagging is used to describe content. This approach to metadata presumably creates more intuitive searches as tags are generally written in the language of the user and it is self-adjusting as these tags will reflect ongoing linguistic trends like new slang.

Not all these features may be emulated by a web 2.0. endeavor and in fact OôReilly concedes that excellence in one area may be a greater qualification than the presence of all these components. Anderson (2007) identifies six ñbig ideasò behind web 2.0 which comprise individual production and user generated content, the harnessing the power of the crowd which is a refinement of Oôreillyôs concept of ñcollective intelligenceò, data on an epic scale which is equivalent to Oôreillyôs re-mixable data source and data transformations, the architecture of participation, network effects, power laws and long tail and open-ness. He seems to take a less technological approach and pursues a rather more social-economic if not political interpretation to the idea than Tim Oôreilly mentioned earlier but there are certain recurring themes notably the idea that web 2.0 as a concept revolves around data and the idea of decentralizing the authorship of internet content. Of particular interest here are data, harnessing the power of the crowd, content and network effects, power laws and long tail.

## *4.2. Justification for a Web 2.0. Approach for Small and Medium Sized Enterprises*

### 4.2.1. Ongoing Organizational Change

Traditionally the requirements analysis is intended to derive the set of functionality required for the business. In other words the goal is to satisfy the client. Whereas satisfying the client is paramount limiting the product to a small set of requirements as described by the client is problematic in that the organization is a changing entity. Chaos theory of the organization stipulates that the organization is never the same at any two times and that the organization has ñstrangeò attractors or rather tends to certain equilibria but is never static thus requirements are bound to change. As a result the solution that results from a given requirements analysis is quickly outdated as the organization changes thus the solution must also change. Therefore a model of the organization is required within which these changes can be interpreted. In this work I employ a business process view of the organization. In the next section I analyze the changes that may occur within the framework of the organizational life cycle and from these derive quality concerns.

### 4.2.2. The Small Business Myth and Organizational Growth and Development

The organizationôs ambition is to grow.  According to Quinn and Cameron (1983 organizational growth takes place in eight dimensions namely size, bureaucracy, division of labor, centralization, formalization, administrative intensity, internal systems and lateral teams and task forces for coordination. As the organization matures its scope increases in terms of these dimensions, that is it gets larger, more bureaucratic, more centralized, bureaucratic and this also applies to division of labor, formalization all the way through to lateral teams and task forces for coordination. This growth occurs within a larger set of events that combined comprise the organizational life cycle. The organizational life cycle is a model that depicts the growth and maturity of the organization and it is thought that the organization goes through a series of predictable phases akin to those of a living organism from inception or birth to maturity and finally decline (figure 6). Given an organization is complex enough it may have several growth curves each relating to its several business ventures and therefore curves may overlap and many a time they may contradict leading to confusion (Steffen, 2001) concomitantly the organization may exhibit growth spurts (figure 5) at any phase altering the shape of the curve.  As Withey (2002) points out organizational development does not necessary strictly conform to a smooth curve but cataclysmic events can occur that may alter the shape of that curve forever.

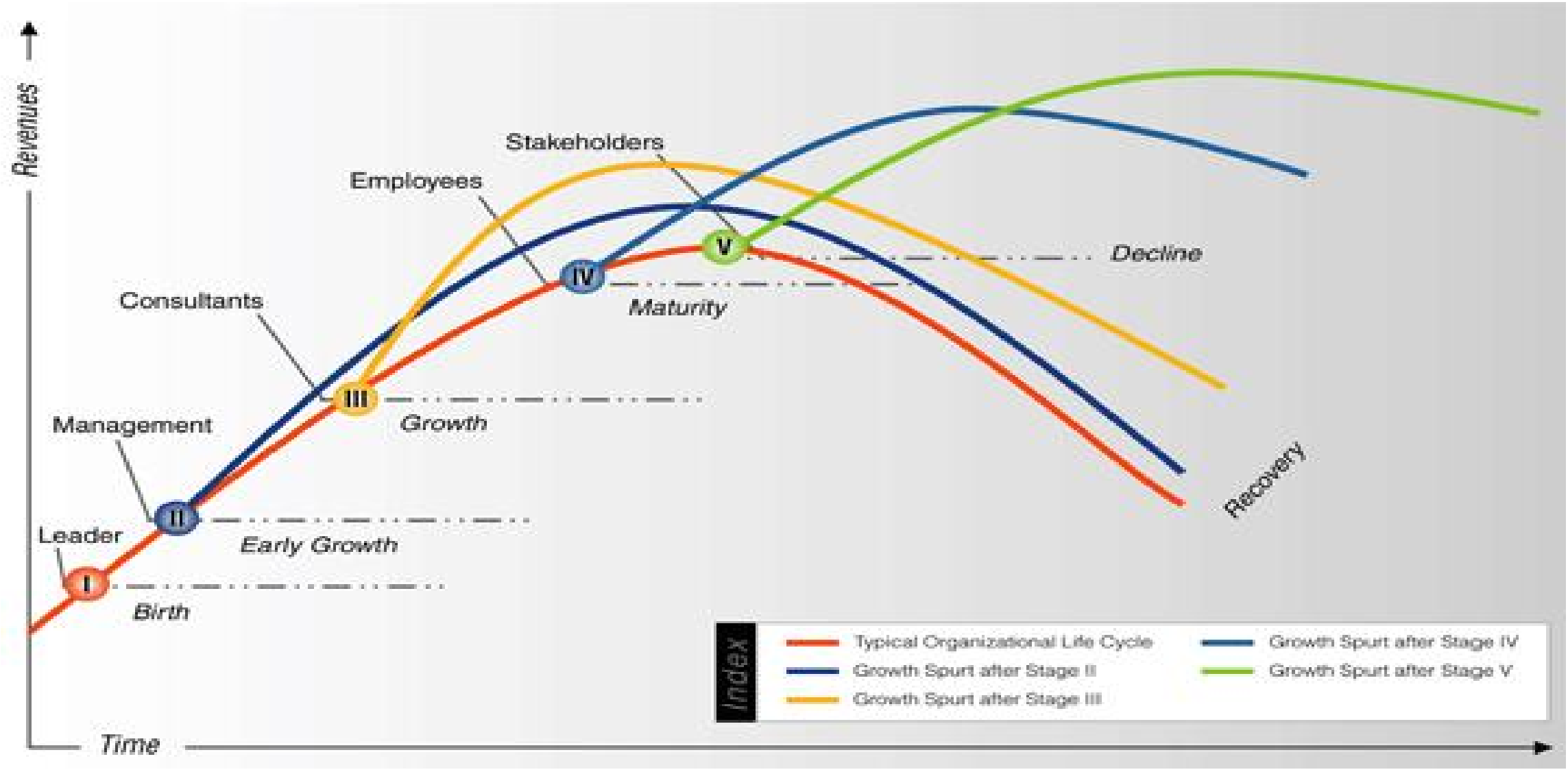

*Figure 5: Graph depicting organizational Life Cycle with overlapping phases*

*(adopted from http://www.iveybusinessjournal.com/UserFiles/Image/nov_dec_2007/sampath3.png)*

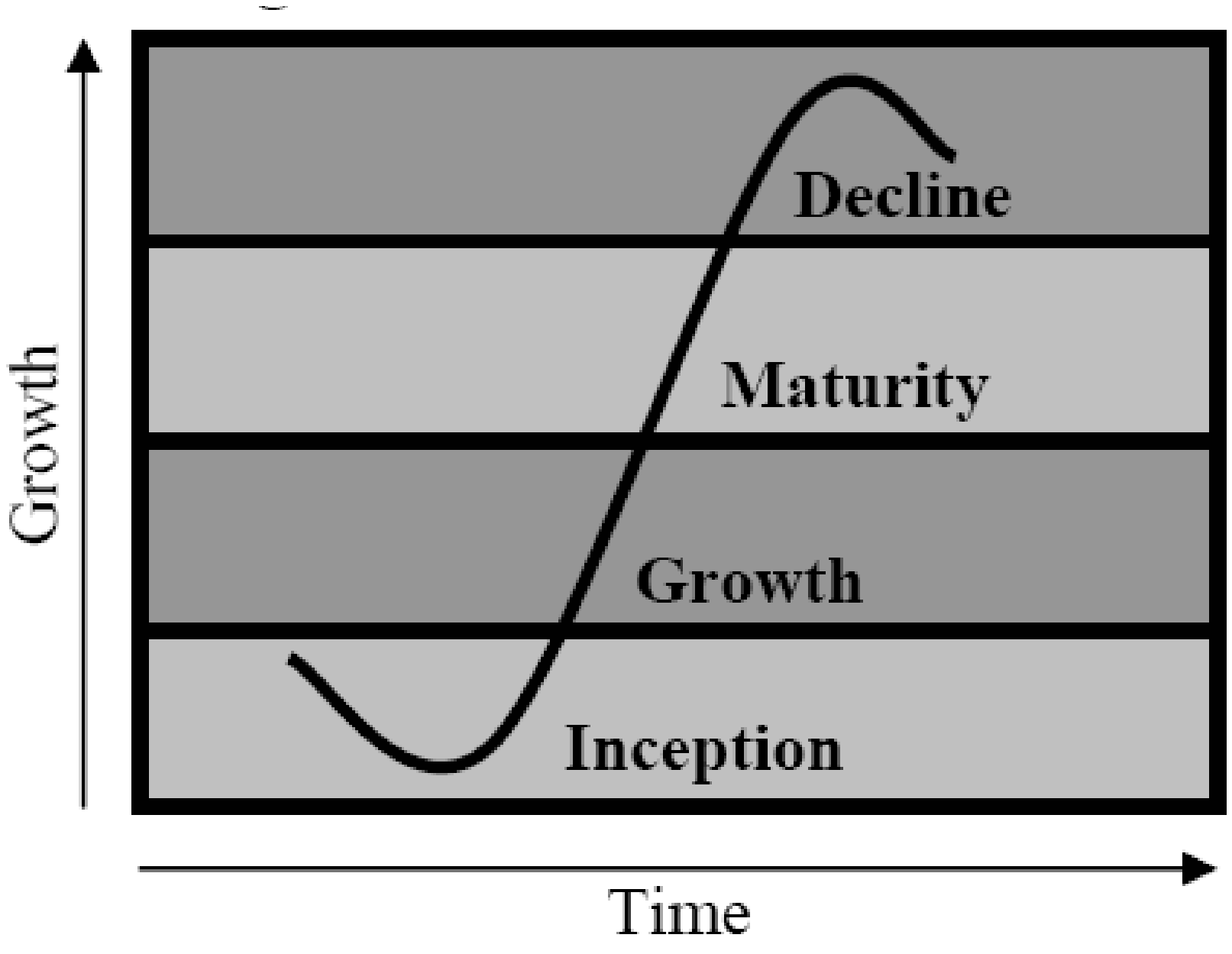

*Figure 6: Chart depicting sigmoid organizational growth curve (adopted from Steffen (2001))*

The small business can be conceptualized as an organization in the inception or ñstart upò phase of the curve (Tyebjee, Bruno, & Mcintyre, 1983). The strategic relevance of this in terms of ICT is that it gives an insight into the level of complexity and definition of the organizations internal processes as the practice has been that prior to designing a solution or strategy for a given client the business processes are analyzed as ICT is normally employed as a tool to rationalize and optimize the organizations business processes. In many cases these processes are directly mapped on to the solution design via business rules that themselves are mapped on to the programming logic. Thus the growth and complexity of ICT initiatives tends to mimic that of the organization. The impact of cataclysmic events is most profound on the small organization characterized by low capital, simple processes and low market share. Should the organization survive this period of uncertainty and stabilize then it enters an exponential growth phase but before that an initial dip is typical as is evident on the sigmoid curve representing brief upheavals as the organization adjusts to an adverse environment. The shift from an ñembryonicò organization to a ñgrowingò organization occurs quite rapidly and it requires immense management effort and vision for the organization to react to these changes. The shift from infancy to maturity is not smooth though and is often punctuated (Withey, 2002). The organization will ultimately derive its longevity from its ability to satisfy customers and expand its market share. This ability in turn is a direct reflection of its internal competencies.

The huge challenges to ICT is having the flexibility and scalability to match the changing requirements of a growing organization. However, given the busty, punctuated and exponential characteristics of organizational growth it implies that this model of the enterprise and such the solution can quickly become outdated. The shift from a few simple business processes to many, complex processes may occur quickly. Additionally each phase or sub phase requires a change in approach strategically which ideally should translate to a change in operations and inevitably a systems change. Overall the complexity of ICT tends to increase with the complexity of the organization for example Klemens and Scott (2000) provide a phased model of Internet presence comprising six phases from a simple web page to a futuristic global grid and the general contention is that these phases correspond to an organization that is increasingly complex. However this increase in size and complexity does not occur smoothly, the organization goes through several upheavals like restructuring and changes in direction thus given this reality it is prudential to impart flexibility and scalability into the organizational ICT infrastructure otherwise ICT will lag organizational growth and as a result of this mismatch competitive advantage may be lost.

### 4.2.3. Quality Concerns:

As a result of this therefore, before the organization's ICT is mature it undergoes a series of re-architecture as the organization's needs change. Overall the total cost of ownership of the entire systems from its inception as a simple web page to a fully-fledged enterprise wide portal that integrates suppliers and clients could be enormous. Additionally enormous risk exists in periods of change. These issues emerge as a vagary of a design process which from a business process point of view endeavors to emulate the organization's process meaning that should any significant organizational reshuffle occur a large amount of systems re-architecture must ensue. This wouldn't have been a problem given the organizational growth strictly conforms to a smooth sigmoid curve and change is predictable but, especially for small young endeavors, change for example growth may occur rapidly as is depicted by the steep portion of the growth curve (figure 5) and in reality some phases of the growth curve will be extremely short or omitted thus a mismatch between ICT and the organization's requirements can emerge very quickly.

Figure 6 below depicts the cost of change against time. This graph has typically been employed to argue the case of continual testing as the later errors are discovered the more costly they are to correct. However, inevitable changes in requirements are not too dissimilar to errors and these undoubtedly will occur much later in the development of the solution, many after words so in this case I employ it as an argument for a flexible systems architecture which minimizes the need for re-architecture.

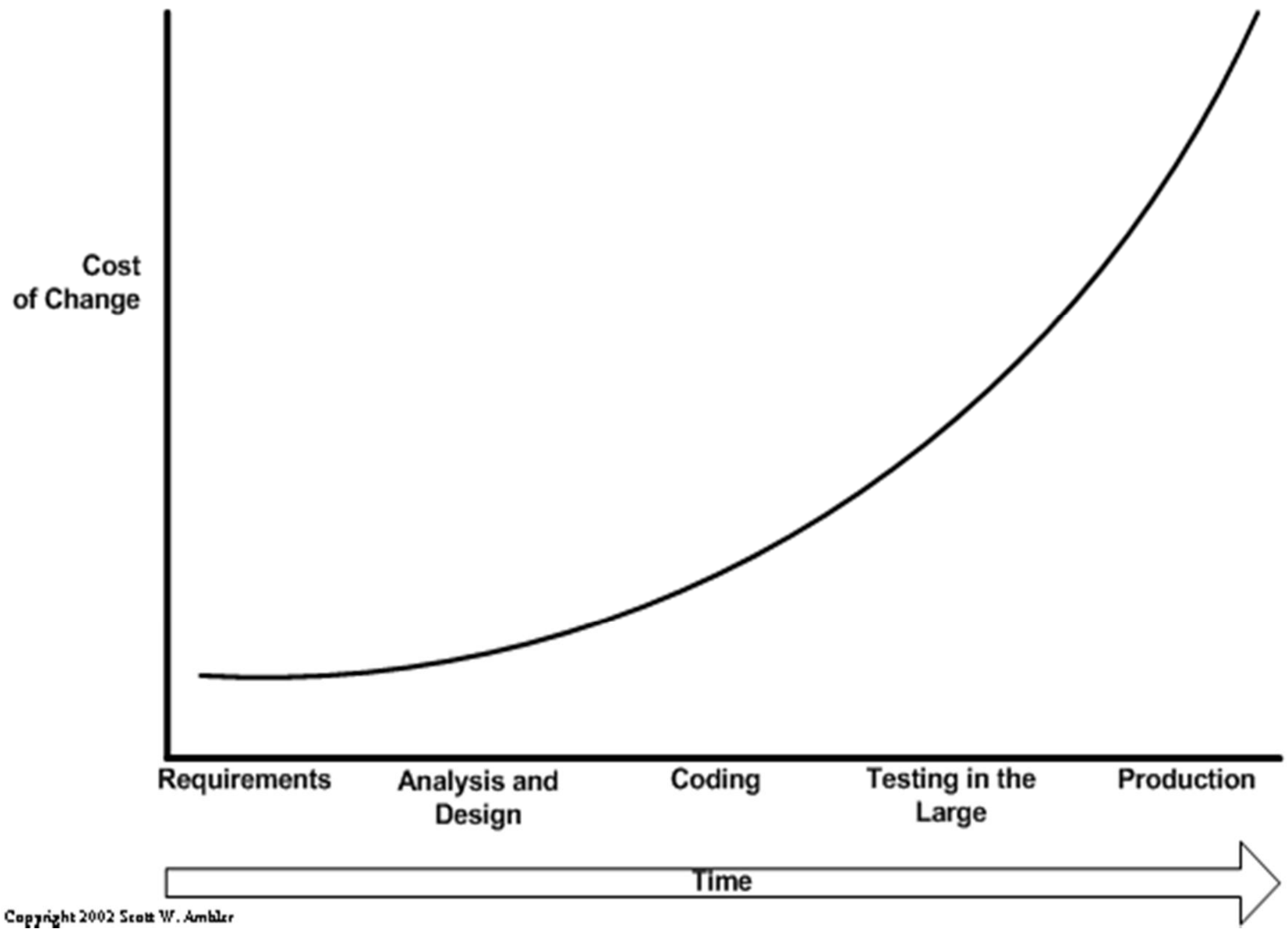

*Figure 7: Cost of Change against time*

*(Source: http://www.agilemodeling.com/essays/costOfChange.htm)*

Itôs also an argument for a more sensitive managerial process. Given the threshold at which the organization recognizes and corrects an error condition is too high (section 3.1.2), thereôs the likelihood that losses incurred will be so great. It is important that the organization as a whole is sensitive to small deviations from desired results and for organizations embracing the web, web 2.0 gives them this extra sensitivity by permitting a higher level of sensitivity allowing the organization to react to a singular individualôs needs without the need to first aggregate information by giving users the ability to generate their own value.

# Chapter 5 ï Web 2.0 in the wider Context

### *5.4.1. Overview*

Web 2.0 has got wider implications than simply revolutionizing the way in which parties interact on the web. Thereôs a host of emergent issues emanating from its impact on the behavior of parties and the technological changes driving it as well as business drivers shaping its development

### *5.2. Business to Business (B2B) Web 2.0*

So far the discussion has mainly focused on the application of web 2.0 to the dialog between businesses and customers but it should be made clear that web 2.0 concepts are applicable to all manner of interactions on the internet. So, the same ideas discussed here are directly transferrable to business to business interactions like partner relationship management. Now more than ever businesses need each other and these strategic partnerships enable businesses to concentrate on what they do best whereas allowing them to benefit from somewhat unrelated establishments by enhancing their capabilities with their services. An ecommerce website can be architected as a ñfacebook connectò website allowing its users to publish blogs, reviews and multimedia via the ñshareò utility therefore exploiting the marketing power of individuals as well as gaining access to an online community estimated at two hundred million (Associated Press, 2009).

Likewise Google.com can benefit from small companies offering niche services by enabling them to use their maps utility. Google benefits by achieving higher visibility whereas the smaller company would enrich its users experience with the advanced capabilities of a powerful geographical mapping system it would also have access to a growing user base some of whom may not even be aware that they are using Google and the intelligent algorithm employed by Google would aggregate user information from this large pool of users and effectively increase in intelligence to provide a more intuitive service. This benefit is distributed across all Googleôs partners. Another example is paypal.com. Any website can implement a secure payment gateway by embedding Paypal.com into its shopping cart. So, a small business would benefit from advanced encryption services and Paypal.com gains higher visibility. This not only shortens the development time for new e-commerce ventures but also allows non technically oriented businesses to concentrate on their core competencies rather than worry about the technical implementation of their website features like encryption, behavior analytics as more and more these features are being packaged into services that can be integrated over traditional protocols like Hypertext Transfer Protocol (http), Secure Socket Layer (SSL) and Internet Protocol (IP). Recently XML (Extensible Markup Language) has been embraced as a medium for the standardization of internet communications as the web shifts towards open platforms. The value for small business is that they can now provide unprecedented levels of functionality in a fairly short period by integrating disparate services into theirs to create a rich user experience.

### *5.3. Driving the Paradigm shift*

All this emphasis on collaboration and democratization of the internet is inevitably changing the entire face of doing business on the web. The first period of internet growth (web 1.0) was about closed platforms and businesses applied a sort of technological protectionism by attempting to own their services and it worked for a while but as the popularity of the internet has increased to a probable one point six billion users worldwide which is approximately a quarter of the world population (internetworldstats.com, 2009) it has accordingly attracted intense business interest on its way there and the competition on the net is extremely fierce. Additionally the continual onslaught of the open source movement has further fragmented that market place. Therefore marketing has emerged as a strong decider of internet business success as companies have been observed to rise and collapse regardless of their technological expertise. So, in the era of marketing it follows that since the web has no physical existence and exists merely as a loose electronic database of interconnected documents then any online marketing effort would have to be implemented on this electronic media. So, whereas technological excellence is still a driving force behind the dominance of companies like youtube.com, and to an extent a level of protectionism is necessary for companies like Google.com whose search algorithm and load balancing techniques are an integral part of its business success there has been a positive move towards more open platforms. By this I mean companies have acknowledged the need for collaboration and have elaborated interfaces for connecting with their businesses as a way of facilitating collaboration. Technology is still important but more and more it is the technology of collaboration that is the architectural core of web 2.0.

So, the ownership of services is decentralized in a way although the internal functioning of the service is obscured from its users. Take for instance any business can implement the Google API (Application User Interface) on its website and create a unique location aware service so in principle the ownership of that service would be transferred to the business although the core technology that defines that service and the ultimate responsibility for keeping it running lies with Google.com. However the move to ñopennessò itself requires first of all a move towards platforms as the preferred architectural strategy for web 2.0. As a consequence huge emphasis has been put into addressing the interoperability problem with new standards being proposed the most influential of which is, as earlier mentioned, XML. Again here the value of this for small businesses is that if they can implement a platform, however primitive, they would have effectively provided the tools for an incremental developmental approach in which anyone could be act as a freelance web developer for the company and add to the overall functionality of the system.

### *5.4. The Other side of the coin: Securing Web 2.0.*

Its benefits aside, the proliferation of new technologies in varying levels of maturity combined with quick time to market and the higher levels of integration also poses new security challenges. The Secure Enterprise 2.0. Forum (2009) also additionally notes that despite the fact that many of these threats are the result of weaknesses inherent within web 2.0 technologies such as wikis and AJAX many of them emerge from the new usage patterns characteristic of web 2.0 such as user generated content, mash ups and web services, consumer and enterprise world convergence, the diversity of client systems and increasing systems complexity for example as is depicted by the asynchronous request where a system behavior may be automatically triggered by the server without an explicit request from the client. Specific attacks associated with web 2.0 include:

### 5.4.1. Flattening of the user hierarchy

In the web 1.0 parlance it is assumed there is a hierarchy of system users with administrators at the top of that hierarchy with super permissions and users at the bottom of that hierarchy with limited access that typically involves read permissions but certain web 2.0 implementations like wikis effectively flatten the hierarchy therefore there's considerable risk that security can be breached. Given this scenario is combined with weak passwords and insufficient brute force controls the risk becomes magnified. The Web Hacking Incidents Database (WHID, 2009) reports that on the fifth of January 2009 twitter accounts of famous people including U.S. president Barack Obama were hacked into by applying a brute force attack on an administrative account. One big architectural concern stems from the issue of a single sign on, that is, a direct consequence of the reconfiguration of the web as a collection of remix able services meaning by signing on to a single website users may gain access to all contributory systems therefore just in the same way all these systems enjoy the benefit of the synergies that result from combining their capabilities they would also inherit each otherȎs security weaknesses.

### 5.4.2. XSS and CSRF

Other specific vulnerabilities include stored XSS (Cross Site Scripting). In this type of attack an attacker takes advantage of a site that enables a user to publish formatted information like HTML to embed dangerous code. In one variant of the attack the attacker embeds a web page containing malicious code in an HTML frame and employing a more powerful script like JavaScript obtains an unwitting users information. To the user on the target system everything appears normal but unbeknownst to them their information is being collected. Christey, Steve and Martin (2007) remarkably report that XSS has become the predominant form of system attack overtaking buffer over flows. WHID (2008) reports that hackers were able to gain access to yahoo session ids by injecting concealed JavaScript allowing them redirect users to an external server. The ids were yahoo wide ids meaning they could gain access to all yahoo services including yahoo mail. Cross Site Request Forgery (CSRF) on the other hand involves a user visiting a malicious site and then in much the same way XSS is effected the attacker obtains a session cookie to a system which the user is already logged on and issue commands on the user's behalf. This vulnerability mainly stems from the use of AJAX as opposed to older systems a client request may not generate user feedback therefore making the attack invisible to the user. This type of attack is limited by "same origin policy" which is a security policy implemented on the client side that allows scripts originating from the same web site to access each otherȎs objects, functions and methods but not those from different sites but the pervasive nature of web 2.0 creates a problem in that these policies may not be uniformly implemented across the wide spectrum of devices and platforms on which web 2.0 operates.

### 5.4.3. XML, XPath and JSON injection flaws

Injection flaws are another area of concern. The participatory nature of web 2.0 introduces new vulnerabilities such as XML, XPath and JSON injection attacks. With XML injection a user gains the ability to modify XML tags as a result of poor authentication procedures such as the application not blocking reserved word or special characters used in XML thereby allowing the user to not only alter the content but the logical structure of the document as well. Using XPath injection an attacker may gain access with the XML database completely bypassing the application all together. JSON (JavaScript Object Notation) like XML is a text based, human readable data exchange format and is primarily used as an alternative to XML in Ajax programming. The problem is JSON is syntactically correct JavaScript therefore it can be interpreted by any browser and effectively a malicious script can create JSON objects on a target system simply by a user visiting a malicious website. Papadimoulis (2008) documents how he was able to bypass security for The Federal Supplier's Guide website. The major flaw emanated from the fact that password validation was done with a client side JavaScript function and the password was global. The attacker in this case was led to suspect that the website employed client side authentication because of the speed at which the error was returned on entering the wrong password and out of curiosity checked the page code which revealed the javascript. The password authentication script looked like this

```
2  <script language = "text'javascript">
3  <!-- This code allows people to enter by using a form that asks for a UserID and password
4  function pasuser(form){
5  if(form.id.value=="buyers"){
6          if(form.pass.value=="gov1996"){
7          location="http:officers.federalsuppliers.com agents.html"
8  }else{
9  Alert("Invalid password!")
10 }
11 }else{Alert("invalid UserID")}
12 }
13 -->
14 </script>
```

Therefore accessing the password protected resource was as simple as copying the url (line 7) and pasting it in the browserôs address bar. The simplicity of the hack was almost comical but on the other hand it clearly depicts how a system can easily inherit a weakness from another system with which it is sharing information. The resource in this case was meant to be for suppliers only but as demonstrated by this hack it was more or less open to the general public. The fact that the password was global meant that his session parameters were valid for this resource.

# Chapter 6 ï Case Study and Practical Application:

## *6.1. Overview*

From a development perspective web 2.0 may be interpreted as a form of Rapid Applications Development (RAD) approach. The idea of having immensely powerful platforms with open APIs through which functionality can be rapidly deployed through REST (Representational State Transfer) is in a way an offshoot of the RAD concept. However, in this section I wish to engage with web 2.0 at the level of its ideological influence on the development process. The initial version of the web, assumed a more or less architecturally static web with well-defined slow changing components. It relied on a formal requirements specification process and was heavily invested in the elimination of ambiguity between the intended users and the developer in pursuit of a singular view of system requirements. However, with web 2.0 there has been a progressive departure from this idea and an acknowledgement that at any given time thereõs a multiplicity of views between development and other stakeholders particularly users as differences in opinion between the two parties can never be totally reconciled and more importantly an acknowledgement of the fact that all these potentially divergent views are necessary and that they can coexist. Therefore, at the heart of the web 2.0 development process is the provision of the tools for the coexistence of these many probable views. Whereas the old concepts of software development and project management do apply web 2.0 concepts most profoundly exert their influence in the planning phase and the implementation phase.
So, at a time when the debate on software development approaches is far from resolved web 2.0 promises even more heated discourse although by the nature of its current implementations its probably an argument for agile systems methodologies. Additionally the system is never complete but constantly being and rebuilt at different levels of the architecture.

Here I derive a framework for the practical implementation of web 2.0. I intend it as blue print upon which any methodology may be based. The approach I chose for this case is an incremental development process as a result of the acknowledgement that the company may lack the skills and (or) the funds to implement an authentic web 2.0. solution so it is assumed that as the company grows its interactions will inevitably increase in number and complexity and this will drive the maturation of its web 2.0 effort. Appendices 1 through to 4 are therefore based on the first increment of the solution also referred to as the core product which is the template upon which subsequent versions of the solution are based.

As posited earlier the problem lies in moving from the descriptive theoretical discussions to a prescriptive implementation process. At the moment the development and application of web 2.0 has been haphazard and a coherent framework for the development and deployment of web 2.0 is lacking. A great deal of emphasis has instead been dedicated to the technological angle of the movement as proponents seek a tangible justification of the terminology. With a myriad of development techniques and methodologies to chose from as a legacy of the unresolved web 1.0 debate on the best development approach it is important to define a framework within which the implementation effort can be managed and disciplined. Such a framework would have to meet the following requirements:

- Technology Independence: A good development framework ought to be independent of technology as by its nature web 2.0 features a broad spectrum of technologies
- Theoretically Justifiable: A good development framework ought to be well grounded in theory that is it must make sense
- Practical: A good development framework ought to be prescriptive enough to be easily translatable to a practical process.

## *6.2. A Propositional Web 2.0 Development Framework*

As web 2.0 is driven by the need to enhance collaboration between web participants it is logical that as a starting point for any web 2.0 endeavor the developer identifies those participants. Then the development process should define ongoing dialog between participants by determining the information flows that shape their interaction. The next step is followed by defining the parameters of the participantsô ongoing dialog in terms of interfaces between participants. This phase involves defining APIs and making choices about which technologies to implement. The final phase involves the implementation of the solution. In actuality this process is super imposed upon the ongoing life cycle based software development process therefore it doesnôt make the life cycle based methodology of web 1.0 redundant but rather improves it. Figure6 below depicts the two processes occurring concurrently as they should. Time is assumed to progress downwards and as a consequence of the dynamic nature of web 2.0 development is a continuous process as system capabilities are added or subtracted on demand.

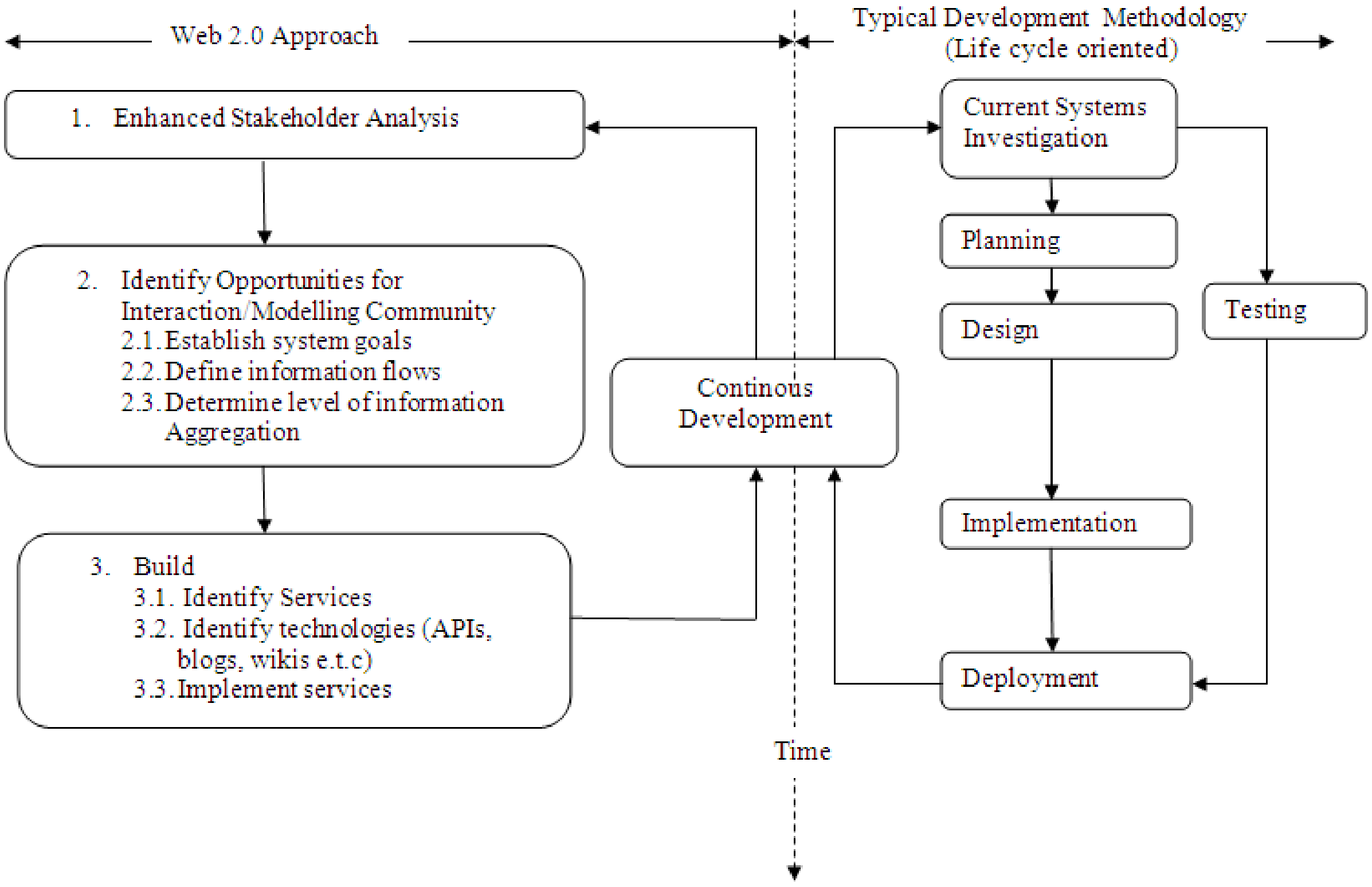

*Figure 8: A Propositional web 2.0. Development Frame Work*

## *6.3. From Theory to Implementation*

### 6.3.1. Organizational Background

At this point I do feel itȏs important to give a brief overview of the client organization. Keymaid is an ambitious startup in the potentially lucrative sanitation industry. As a new company it has to quickly establish itself within the industry by employing a thorough and comprehensive marketing strategy and it must quickly gain customers to extend its market exposure and at the same time increase market share through customer retention. Critical to these ambitions is the idea of customer satisfaction. The proposed solution employs the wide reach of the internet and the principles of Web 2.0 to empower participants.

### 6.3.2. Application of the Framework

***Multiplexing and Polymorphism***

Before I delve into an explanation of my propositional framework, I wish first of all to introduce to introduce two core ideas that foundation of my framework. Polymorphism is the acknowledgement of the fact that participants exist in different forms (polymorphs). It is important with web 2.0 implementations because participants (the sources and receivers in the communication circuits) exist in a many forms and exhibit different behaviors and characteristics in these forms therefore to fully investigate the interactions that may occur it is important to distinguish between different forms of the same participant. The customer may exist on a mobile device, a stationary PC terminal, may be online or offline and maybe online but on another site all together. I propose four broad categories (base states) with in which polymorphism can occur. These categories correspond to different user states and comprise online user, offline user, mobile user and stationary user. Being cognizant of these states is important in recognizing all the possible opportunities. It is more useful to view the constituent states of a polymorphic source as different sources all together therefore a distinction is made between an online user and the same user in an offline state. Still from these broad categorizations further polymorphism can occur from which more communication circuits can be implemented and new information flows enabled. Figure 9 depicts polymorphism on an Entity Relationship Diagram as a form of specialization

Multiplexing on the other hand is the acknowledgement of the fact that these polymorphs can implement multiple information flows on a single communication circuit. This idea is important when making design decisions about the number of views (information flows) you wish to implement. This is to do with the richness or depth of data that you intend to accord to your participants. The more the detail the more information flows you actually implement for a given communication path. As with when micro formats (see section 4.14) are used ï at a logical level a single http conversation can be dissected into a hCard channel, a hCalendar channel, a hAudio channel. This helps the designer to set important goals for the final system functionality from a very early point in the application development

| | |
|---|---|
| ***Online***<br>• Data synchronization<br>• Streaming options<br>• Strategic partnerships | ***Offline***<br>• Asynchronous communications<br>• Downloaded content |
| ***Mobile***<br>• Power limitations<br>• Small display size<br>• Security concerns<br>• Memory limitations | ***Stationary***<br>• Large display size<br>• Unlimited power<br>• More secure<br>• More memory |

*Table 1: Table depicting polymorphism with base states*

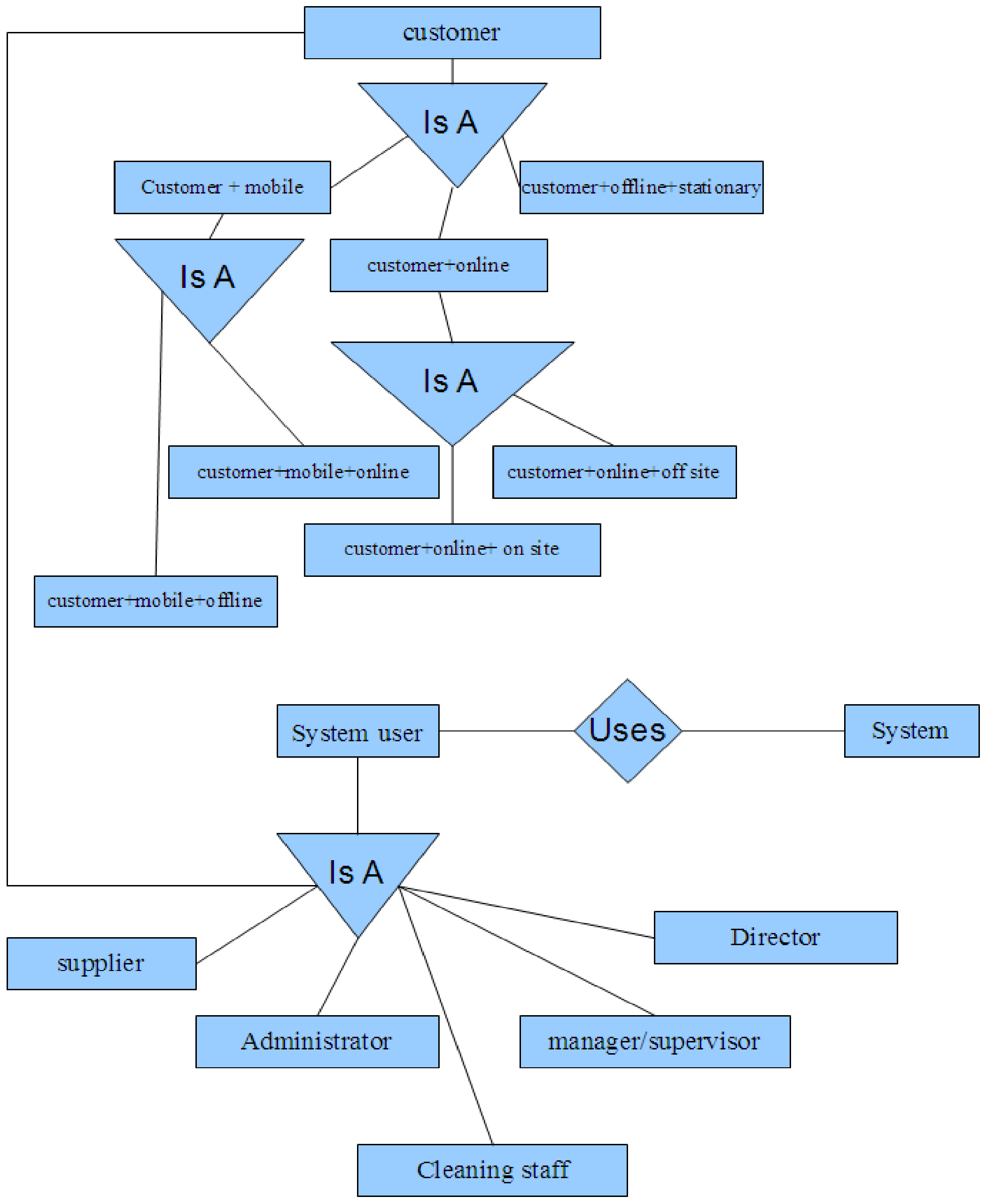

*Figure 9: An ERD adaptation depicting polymorphism as a form of specialization*

### *Step1. Enhanced Stakeholder Analysis*

Stakeholders here are all those parties that affect the system or are affected by the system. In other words they are the originators and consumers of data and services. The table below outlines these participants along with their interests and opinions as far as the system is concerned. The last column defines the information generated by each stake holder. It is important to note that this information possibly translates to huge quantities of data whose meaning may not be obvious till it is aggregated although as in the case of business communications it can be quite explicit. It also goes to demonstrate the view that here as opposed to web 1.0 participants are viewed as sources of information rather than simply as consumers of information. They are active agents capable of marketing and product improvement and even modifying the structure of the system through their interaction with it. The typical web 1.0. viewpoint which is exceedingly biased towards identifying user requirements as users are viewed in a passive sense and the emphasis is on serving them and little room is given to the idea that they can actually be servers of information. Another important idea proposed here is that of polymorphism which is explained shortly. Using polymorphism the traditional stakeholder analysis is ñenhancedò.

The idea that web 2.0 stakeholders are not merely passive recipients of information from a singular authority is dealt with simply by slightly modifying the stakeholder analysis process. Table 2 is a stakeholder analysis for Keymaid Cleaning Agency with a column for the information requirements as well the information generated by each user state. This allows the shape of the dialog to be visualized early on as the probable conversations/ information paths are simply derived by matching a given participantôs information needs with another participantôs output information flows.

| Stakeholder | State | Major Value | Major Considerations | Relevant Information Flows | |
|---|---|---|---|---|---|
| | | | | Information Requirements | Information Generated |
| Directors | | • improved employee productivity;<br>• cost savings, planning and coordination functions | • Interface usability<br>• Data summarization (High level of aggregation) | • Financial Data<br>• Operational details<br>• Sales information<br>• Supplier information<br>• Market data<br>• Human Resource Information<br>• Procurement and Logistics | • Customer greetings<br>• Service updates<br>• General thanks<br>• Resource tags |
| Cleaning staff | | • Lower travel expenses<br>• Higher job flexibility<br>• Fair automated job assessment | • Interface usability<br>• Assessment Algorithm | • Job rotors<br>• Impromptu Announcements | • Complaints<br>• Suggestions<br>• Nomination of peers for awards<br>• Resource Tags<br>• Marketing Recommendations |
| Managers/ Supervisors | | • Scheduling functions<br>• Job rotors | • Interface usability | job preservation | • Employee ratings<br>• Memos<br>• General Thanks |
| Customers | | | | | |
| | Online + onsite | Convenience, time saving | • Interface usability<br>• Information content | • Service/ product descriptions<br>• Price information<br>• Company information<br>• Contact Details<br>• Help manuals | • Service ratings<br>• Complaints<br>• Marketing Recommendations<br>• Resource Tags<br>• Enquiries<br>• Orders<br>• Cancellations |
| | Online + offsite | | • Interface usability | • Reviews<br>• Advertisements | • Comments<br>• Reviews<br>• Advertisements |
| | Offline+stationary | | • Company visibility | • Downloadable/printable | • Tracking cookies |
| | | | | Content:<br>➢ Job history<br>➢ Payment data<br>➢ Terms of Reference<br>➢ Privacy policy<br>➢ Order details summaries | |
| | Mobile + logged in | • Customer convenience<br>• Company visibility | • Security<br>• Connectivity<br>• Company visibility | • Job status<br>• Job bookings | • Orders<br>• Cancellations |
| | Mobile + logged out | • Customer convenience<br>• Company visibility | • Interface design | • Downloadable Applications | |
| Suppliers | | • Supplier convenience<br>• Order fulfillment | • Supply chain integration<br>• Interoperability | • Current inventory Levels<br>• Order information | • Stock levels<br>• New products<br>• Ongoing promotions |
| Administrator | | • System stability<br>• Service reliability | • Security | • System usage statistics<br>• Systems configurations | • Systems configurations |

*Table 2: Table showing enhanced stakeholder analysis*

***Step 2: Opportunities for Interaction/ Modeling Community***

These interactions are generally speaking communication circuits therefore this process involves identifying the possible communication circuits or conversations within the community of stakeholders. Shanon (1948) views a communication system as comprising two participants namely the source and the destination who communicate via a medium (figure 10). Shanon also posits that multiplexing can occur over this medium. With multiplexing a single medium can be split into various channels to transmit multiple messages simultaneously. In web 2.0 parlance all participants are potentially information sources or destinations (fig 10) therefore if all the possible circuits are implemented, web 2.0 participants effectively form a mesh like communication system in which every participant is connected to every participant (figure 11). Therefore assuming a fully connected system of n participants the total possible communication circuits N is given by the equation 3below.

$$N = n\,(n-1)/2 \qquad \text{é é é é .3.}$$

However, the most likely actual implementation of this topology is star like (figure 12) with participants communicating via a central database or ring like where participants acquire information indirectly from other participants. These information circuits are implemented via the application. Any information circuit may implement a variety of information flows akin to communicating multiple ideas via a singular conversation.

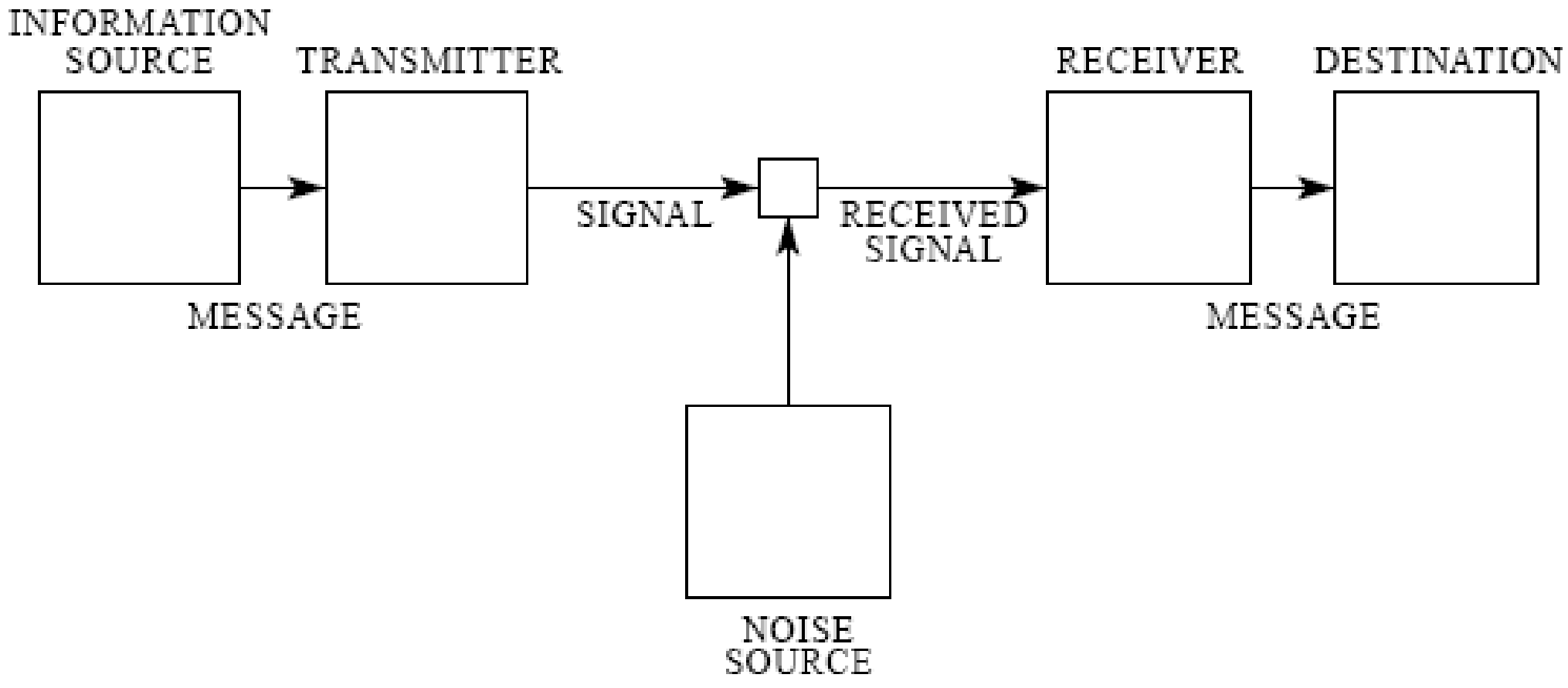

*Figure 10: Schematic Diagram of a General Information System (Shanon, 1948)*

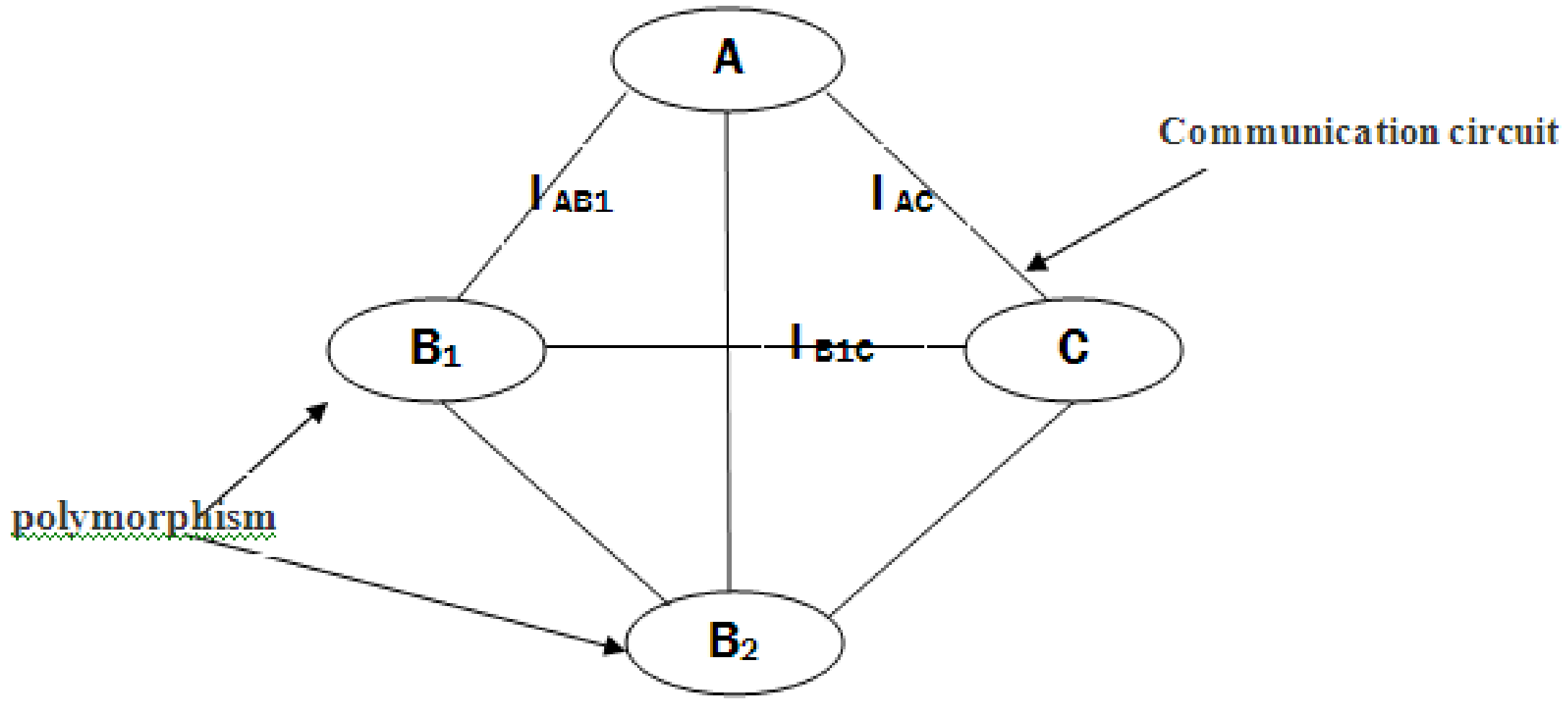

*Figure 11: Schematic Diagram depicting a fully connected community of three participants*

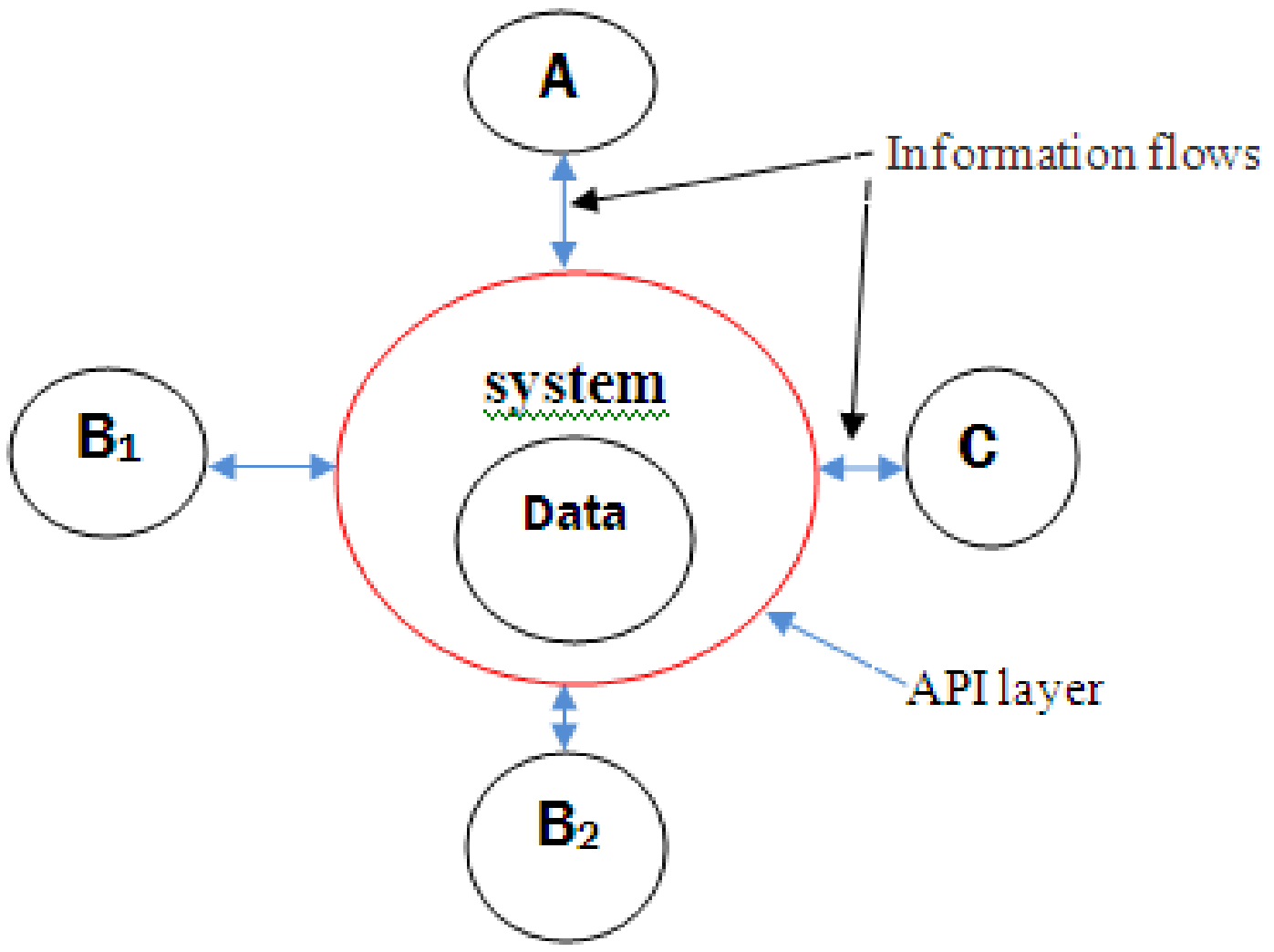

*Figure 12: Schematic Diagram depicting the most likely actual implementation of the schema in fig9*

### *i, System Goals*

A certain cost (c) is incurred with the implementation of each possible information communication circuit and a certain utility (u) is accrued. The net utility (U) of an information flow is the difference between its utility I u and the cost of implementing it I c (equation 4).

$$U = I_u - I_c \ldots\ldots\ldots\ldots\ldots\ldots 4$$

Generally the goal is to create a system such that the benefits derived from it exceed the cost of implementing it at least in the long run but in the short run new system implementations will typically have a negative return on investment (ROI) therefore it is important to determine the position of the business to establish feasible values of I c and weigh development options. Determining which communications circuit to implement is best handled as a collaborative process between the client and the development team. A brain storming session with relevant participants might be sufficient to determine the base functionality of the system

### *ii, Determine Information Flows*

It has already been determined that each participant is a potential source and receiver of information and knowing the number of participants N in any scenario equation 3 gives the possible communication circuits therefore this process is rather more a process of elimination rather than of discovery. By applying an elimination criterion (in this case an economic criterion) we defer the implementation of circuits for which the expected utility is below a justifiable level. However, it generally makes more sense to visualize the total function of the product (× $(I_u - I_c)$ rather than to interpret cost in terms of individual functionalities or information circuits/ flows for this matter a complete design is derived using a normal requirements analysis which is enhanced by the idea of polymorphism and the cost minimization (benefits maximization) would be implemented at the build phase when technological decisions are made. . Figure 13 below is an illustration of the ñcommunityò of keymaid Cleaning Agency. I employ a UML 2.0 communication diagram to illustrate the idea of building community. The difference is that here instead of applying the communication diagram to the software system components I apply it to the system that is the community of stakeholders. Additionally in acknowledgement of the fact that such a system comprises many subsystems the diagram is split into two portions by the dotted line namely subsystem A and subsystem B. The former comprises the dialog that the organization maintains with the system user when they are away from the site in an alternative locale in the internet universe via collaborating APIs whereas subsystem B comprises the dialog maintained by the system and its users whilst the user is onsite. Additionally a distinction is made between these two user states by way of polymorphism which was earlier described. The ongoing ñconversationsò are labeled A or B depending on which subsystem they belong to.

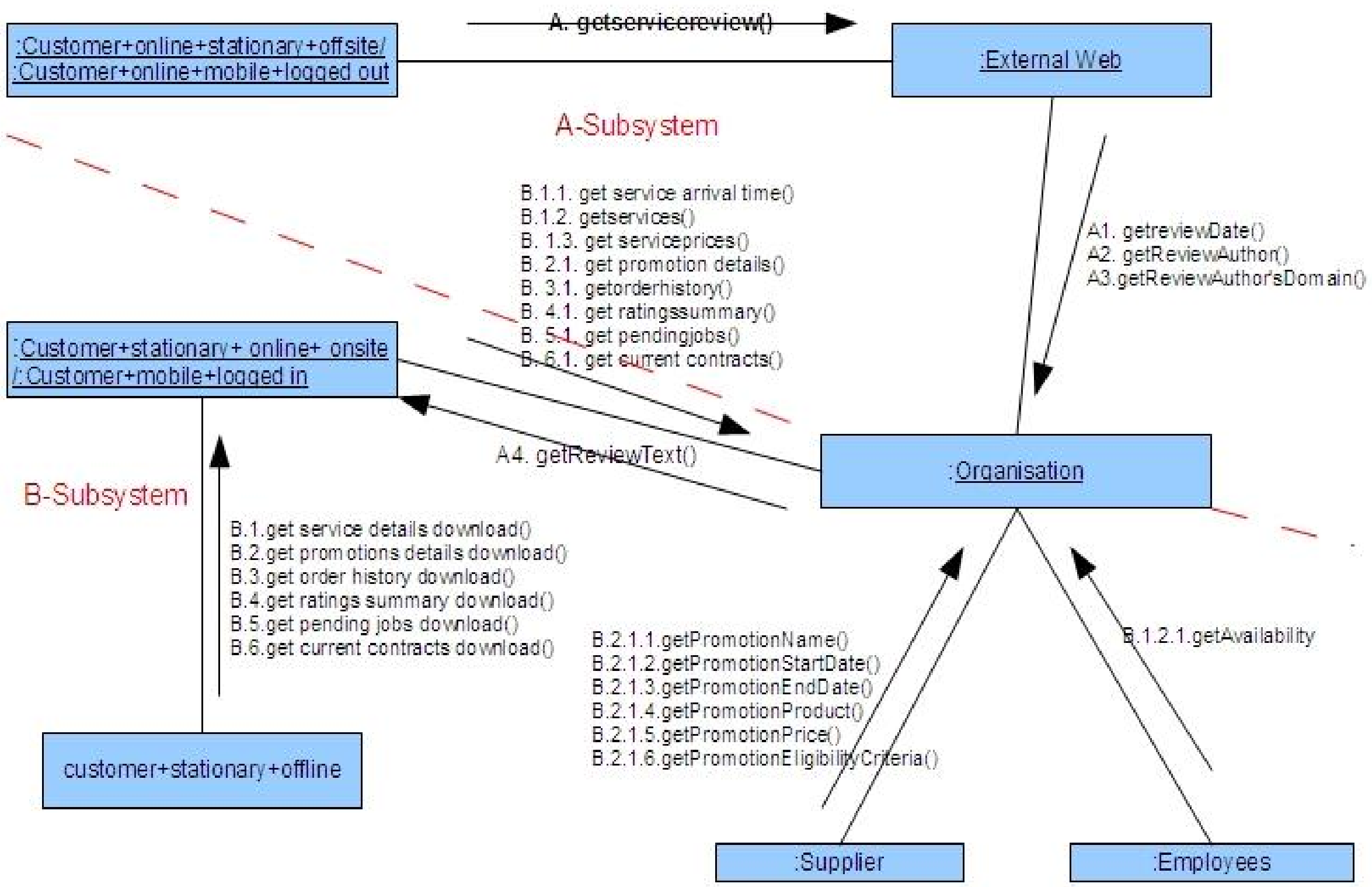

*Figure 13: Modified UML 2.0 communication model of Keymaid's "community"*

## *iii, Power laws:*

The concept of power laws refers to a probability distribution where the probability of a random value being a certain value is proportional to a negative power of that value as expressed in the equation 3 below:

$$P(X = x) \propto cx^{-k} \text{ where } c > 0,\ k > 0 \ldots\ldots\ldots 5$$

Its applicability to web 2.0 is a consequence of the structure of the web which is itself structured as a network of links which can be interpreted as edges within a network and sites as the vertices a power law distribution is observed in which a small number of websites have got a very large number of links to and from them. This implies that a huge proportion of the amount of traffic on the web and therefore the overall value of the internet is derived from a small number of participants. Sheun (2008, page 161) observes that social networking sites express network power laws that are often more extreme than the 80/20 Pareto principle where 80% of the results come from 20% of the people with a critical mass of as low as 3% being capable of triggering exponential growth. Bearing this in mind decisions must be made on giving users the capability to communicate in a variety of ways as well decisions on which external participants (and APIs) to involve. For example a sponsored link on Google.com or gumtree.com is a brilliant strategy for any company seeking higher visibility on the web because of the sheer volume of traffic on Google. Similarly techniques which are geared at encouraging the generation of traffic like blogs which give your website a higher search engine placement are viable strategies for achieving visibility on sites such as Google on the basis of this principle.

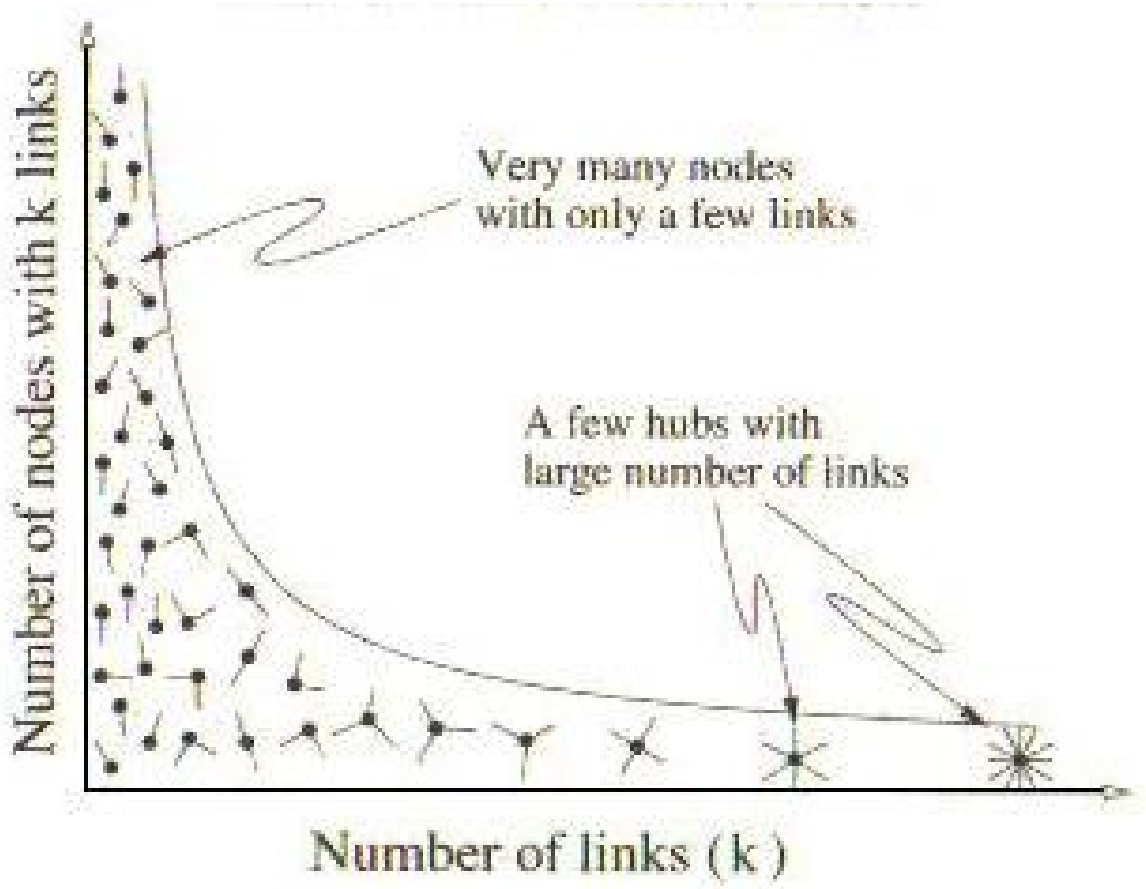

*Figure 14: Curve of nodes against number of links depicting the concept of power laws and long tail*

### *iv, Long tail:*

Looking at the curve thereŝ an area of high frequency, high magnitude events accounting for the bulk of the area under the curve but Anderson (2004) argues that there are situations where the cumulative value of the later part of the curve exceeds or approaches that of the first part (fat tails). Erick et al (2006) demonstrated this for Amazon.com proving that the majority of Amazon.com sales came from obscure, scarcely known titles. In business this particularly holds when certain supply chain efficiencies are achieved so as to lower the costs of distribution and storage which ends can be met by the application of web 2.0. This here is a direct argument for customization as well as product diversification via techniques like Just in Time (JIT) and lean supply chain management and these are better achieved in the interactive domain of web 2.0.

These two mathematical principles (power laws and long tail) form the philosophical basis for web 2.0ŝ economic justification for small business and have direct implications for modeling community and interaction.

## *Determine level of Aggregation*

Again this process is guided by the system goals. Just because there are n (n -1)/2 possible information circuits it doesnt̂ mean they are all worth implementing. It is often not necessary to directly connect participants to each other in a mesh like topology. Take for instance the circuits $I_{B1C}$, $I_{AC}$ and $I_{B1C}$ rather (fig 9) than have $B_1$ interact with C directly through $I_{B1C}$ it can be decided that it does so indirectly via $I_{AC}$ and I

B1C provided it suits the chosen criterion. Likewise the depth to which participants are multiplexed has to be determined. This phase is about determining the richness of communications and how much to slice up information. Certain states may not be relevant for certain implementations and in many cases the base states would generate an adequate number of information flows from which an ample number of opportunities may be derived.

### *Step 3: Build:*

Steps 2 and three are technology independent but this third step involves making technological choices for the implementation of the information flows designed in phase2. The aggregated information flows are packaged into services after which technology choices are made followed by a straight forward process of implementing the information circuits with web 2.0 technologies. Some risk assessment and feasibility analysis will have to be done but it is generally handled by the parallel, inclusive life cycle process. It is at this stage that costs are minimized.

# Chapter 7 ï Evaluation of the project

## *7.1. Objective1*

The first objective was to review the development of web 2.0 and evaluate the current industrial view points on this family of technologies. This was dealt with in chapters 4 and 5 which tackle the history of web 2.0 and its core tenets as well as its impact on the industry as well as hinting on the probable future developments. Both the technological and philosophical dimensions are presented and both views for and against are presented thus the discussion has enough depth as well as breadth. Therefore, this objective is well met.

## *7.2. Objective2*

The second objective was to demonstrate the criticality of web 2.0 technologies through a thorough theoretical review of organizational theory and current technological trends. This objective is dealt with in chapter 2, 3 and 4. Chapter 2 demonstrates the shortcomings of the static web 1.0 model via a systems theory angle, chapter 3 evolves a mathematical model of the management problem and posits that it is in fact a loop whose objective is to maximize quality within certain limits and makes the case for dynamic, flexible platforms as is the case in web 2.0 and effectively posits a financial argument for web 2.0. Finally chapter 4 presents web 2.0 as a quality assurance mechanism and successfully makes the point that in todayô̂s competitive market space web 2.0 is more a necessity than a trend therefore this objective has been well met.

## *7.3. Objective3*

The third objective was to demonstrate the relevance of web 2.0 to small and medium enterprises by identifying opportunities for implementation of web 2.0. This is dealt with explicitly in chapter 4 (subsection 2). The chapter successfully argues that by virtue of the fact that a small organization is essentially an organization in the inception or growth phase of its organizational life cycle a flexible platform is warranted as this phase of organizational development is characterized by instability and rapid change and more often than not cataclysms and growth spurts. Additionally the low revenues associated with new business endeavors and SMEs in general warrant a cost effective solution but the demands of business survival namely quick access to market, customer retention and market growth all require an intuitive and adaptive IT infrastructure and chapter 4 successfully argues that web 2.0 is the most logical approach to combine both functionality and cost saving. Therefore this objective is also well met

## *7.4. Objective4*

The fourth and final objective was to apply the theoretical conclusions to a practical business case to evolve a model for implementation of web 2.0. This is dealt with by chapter 6. The mainly descriptive discussion is compressed into a prescriptive framework to guide the application of these principles to a practical case in this case my sponsor organization. Chapter 6 deals with developing and applying this framework and introduces two core principles namely multiplexing and polymorphism to characterize the dialog between web 2.0 participants. The full details of this process are tackled in the appendices. This objective is also well met

## *7.6. Conclusion*

Overall, I view this project as a success for it has achieved all its objectives within the schedule for the project and both theoretical and practical goals have been achieved.

# Chapter 8 ï Conclusions, Research and Recommendations

## *8.1. Overview*

This chapter presents global conclusions drawn from the overall project and presents the key findings in addition to recommendations as well as going as far as to present areas of further study. The research has been confined to a bibliographic survey and literature search and therefore can only be as reliable as that literature but nonetheless I feel the texts quoted and referred to here paint a representative picture of the current industrial trends and practice in relation to web 2.0 and SMEs.

## *8.2. Conclusions*

### 8.2.1. Bridging the gap

Web 2.0 as a concept is pretty much in its infancy, it means different things to different people and to different establishments therefore success is heavily dependent on the skills of developers and the instincts of organizational strategist. Whereas thereǿs an industry wide desire to implement some of the more ubiquitous elements of web 2.0 like RSS (Really Simple Syndication/ Rich Site Summary) and AJAX programming the challenge as in web 1.0 lies in translating and matching this functionality to the overarching organizational goals and objectives and implementing them within a concise strategy (Hinchcliffe, 2006a). With a plethora of development methodologies and philosophies it is necessary to find a way to streamline a development process, whose methodologies have been inherited from a static model of the web, so as to make it more manageable within a dynamic environment. This creates the need for a developmental framework to bridge the gap between desire and actual implementation. I reiterate that a good frame work is technology independent, easy to implement and theoretically justifiable.

### 8.2.2. The Perpetual Balancing Act

The internet so far has been synonymous with exponential growth and rapid change. From time immemorial there has been a great deal of contributors from different domains otherwise it wouldnǿt have maintained its famed growth. It also was never a perfect system and the speed of change and the characteristic short life spans of dotcoms, the inherent insecurities of TCP/IP as well as immense competition on the web have been extensively written about and debated and web 2.0 doesnǿt solve these age old dilemmas but simply magnifies them. The appeal of web 2.0 as was the case with web 1.0 is made by its business case. It is increasingly being pushed by internet heavy weights like Google.com and Facebook.com who view it as an opportunity to tap into new markets and enhance their visibility on the web. For example Apple has staved off the economic recession on the strength of sales made through its ITunes store which is a prime example of web 2.0 and have defined a comprehensive API for their IPhone enabling them to enjoy revenues from application downloads. Behind these impressive financial figures is a background story of a new generation of millionaires riding on the success of utilities like Google adsense, Youtube channels and Facebook apps. As long as the potential business benefits are compelling enough to the big players who have sufficient funds to define and publish APIs, web 2.0 will thrive.

### 8.2.3. Closed Vs Open Platforms: A Fragile Peace

This is probably one of the most enthralling philosophical feuds of the software world. On one hand the open source movement that agitates for code transparency and in the opposite extreme the closed system proponents who vehemently make the case that software is intellectual property and must be paid for and abstracted from all others. Web 2.0 seems to strike the middle ground. The actual coding and techniques employed by say Google.com are strict trade secrets protected by thick layers of security and encryption but the platform detailing how a remotely hosted CGI (common gateway interface) script can interface with that hidden infrastructure are publically available. It isnøt complete openness but also isnøt absolute closure and for the first time it seems as though both sides can coexist whilst oddly simultaneously claiming victory.

### 8.2.4. A Market for Capability

In web 2.0 parlance capability is the commodity. Thereøs an informal hierarchy with establishments like Google.com, Wikipedia.org and paypal.com all providing capability in terms of data and utilities. These companies are the manufacturers of capability distributing on wholesale and the APIs are like an automated front of house from which the first layer of buyers ñshopò capability. This layer comprises those establishments that integrate these capabilities into their systems via CGI so they are the equivalent of retailers of capability. At the bottom of that hierarchy are the consumers for whom that capability is intended. These capabilities are mostly referred to as services and as a consequence of web 2.0 there is a migration towards a Service Oriented Architecture (SOA) of the web (Hinchcliffe, 2006b) reliant on REST (Representational State Transfer Protocol) rather than SOAP (Simple Object Access Protocol).

This has led to a re-stratification of the web, generating three distinct but often overlapping layers of participants. An example is Facebook.com charging application developers a small fee to place their application in the application directory or Paypal.com charging business accounts a small fee on transactions. Increasingly the emphasis is being placed on providing the widest range of data and utilities to the final users. Therefore the web is migrating towards a subscription model where users can install and uninstall capability or ñapplicationsò for that matter from their accounts. Examples include Hi5.com and Facebook.com.

As a consequence the user has never enjoyed so much utility and power and as much as technology is important understanding the psychology of users is becoming more important and as technology gains more intelligence the influence of user psychology and the social dynamics of online communities is becoming a more important driver of design patterns.

### 8.2.5. The Integration and Interoperability Challenge (I&I)

Web 2.0 by its very nature further underpins the need for interoperability. As a result of this new requirement for greater interconnectedness and standards like XML have emerged to address that need likewise systems integration is strongly justified by web 2.0 as it requires the organizations systems to be more fully interconnected so as to accord users an unprecedented level of interactivity. This challenge is twofold as integration can be approached at the level of technology and also at the operational level. The technology and platforms might have become pervasive and within the reach of small companies but the skills to meaningfully implement it in a competitive environment may still defy the small organization. Hinchcliffe (2006c) reported that despite the fact that ñmashupsò are increasing at a rate of three new ñmashupsò a day their widespread use and deployment was hindered by the integration challenge. Wainewright (2006) provided a comprehensive summary of the major integration and interoperability challenges that face web 2.0, additionally he makes the crucial point that these challenges also existed for the early precursors of web 2.0 in the form of the Internet Business Services Initiative (IBSI).

| | |
|---|---|
| **Integrated access** | • Single sign-on<br>• Account provisioning |
| **Integrated usage** | • Consolidated billing<br>• Common look and feel<br>• Co-ordinating support and SLA responsibilities<br>• Shared reporting metrics |
| **Data integration** | • Key data sharing<br>• Data extraction and backup |
| **Business process integration** | • Workflow integration<br>• Policy integration |

*Table 3: Table Summarizing I&I challenges for web 2.0 (adapted from Wainewright (2006))*

## *8.3. Relationship to Current research*

At the moment there are two main ideological camps with respect to web 2.0. One camp comprises the followers of Tim Barnes Lee the originator of HTML and therefore of the internet who challenge the very relevance of the term and view it as an unnecessary farce and on the other side the proponents of the term and followers of Tim OôReilly who is widely reputed with popularizing the term. Needless to say whilst I appreciate the argument of Tim Barnes Leeôs camp I subscribe to the latter faction. However, the consequence of this debate has been to shape the literature such that proponents of the term have focused more on justifying the term. Therefore most of the work has exceedingly been descriptive (Anderson, 2006; OôReilly, 2005; Berlind, 2006) and as the term gathers steam I feel thereôs a need to translate the huge volume of work to a practical model for the deployment of new web 2.0 endeavors. This is therefore an initial translation of that great body of work into a blue print for web 2.0 implementation.

## 8.4. Areas of Further Study

### 8.4.1. An Appraisal of New Opportunities for SMEs

Thereǿs need to analyze opportunities presented by web 2.0 in relation to SMEs for example the mobile web is an area in which SMEs probably do not have an established presence and as the stationary HTML (Hypertext Markup Language) based internet becomes more and more saturated and competitive the ability for SMEs to migrate towards mobile WML(Wireless Markup Language) based web applications may create considerable competitive advantage. These are opportunities that are made clearer by the idea of polymorphism discussed earlier (section 6.3.2). The market potential of this mobile market has already been demonstrated by the multimedia downloads industry via platforms like ITunes and Sonyǿs online music store. Researchers need to work out a strategy of how SMEs can establish a presence in this emerging market as users spend more and more time on smaller, hand held devices rather than powerful stationary computers.

### 8.4.2. Plugging the holes

The web is an evolving environment and so are the efforts to master it and harness it such as web 2.0. Thereǿs constant change and most of the time it is occurring at such a high speed and with little regulation such that instabilities emerge. One of the more notorious outcomes of such instabilities are security holes. With the fundamental weaknesses of TCP/IP still threatening internet security, web 2.0 has introduced even more security concerns and the old adage that no system is a hundred percent secure has taken on a new significance. It is important to have a thorough appraisal of how to secure web 2.0. This appraisal has to be ongoing as system threats are continuously evolving. Furthermore

### 8.4.3. A Glimpse into the Future: Web 3.0, Web 4.0 and Beyond

The dust of contention is barely settled on web 2.0 and already thereǿs talk of web 3.0. and even web 4.0. Web 3.0 is about an intelligent internet and although the concept at the moment is merely academic its masterminds claim the development world wide web should be perceived in ten year phases (Richards, 2007), the first phase having been web 1.0 followed by web 2.0 and on this basis they foresee a hypothetical future in which thereǿs a progressive shift of emphasis towards the computing back end in the next twenty years and after that a web 4.0 which is a return of emphasis to the front end via extremely intelligent client side technologies. Web 3.0 is also widely referred to as semantic web and although some aspects of it are currently implemented its full definition is largely propositional and futuristic. Tim Barnes Lee likens it to a Scalable Vector Graphic (SVG), in other words access to an enormous information space being guided by a small set of parameters mapped on to a semantic code as a complex graphic may be rendered from a set of points in SVG.

On the basis of these ongoing developments, I conclude this discussion by saying that web 2.0 is a journey, a means to an end and not an end in itself.

# Appendix 1 ï Background, Business Opportunity, and Customer Needs

Keymaid is an ambitious startup in the potentially lucrative sanitation industry. As a new company it has to quickly establish itself within the industry by employing a thorough and comprehensive marketing strategy and it must quickly gain customers to extend its market exposure and at the same time increase market share through customer retention. Critical to these ambitions is the idea of customer satisfaction. The proposed solution employs the wide reach of the internet and the principles of Web 2.0 to empower the customer. The customer from any computer connected to the internet can order a service, pay for it and rate it. These ratings are used to measure the productivity of individual employees and each employee can log in to view their personal ratings in addition to work rotors and payment information. Thus the system is a self regulatory management tool as employees whose performance rating is below average motivate themselves to fit within the group.

Fairness is protected by an appeal utility which allows employees to appeal their rating. The employee rating is compounded from the customerôs rating and the supervisorsô ratings. The system also provides additional capabilities like payroll management, basic financial and accounting functions and inventory management. These are functions that are required by each company but given the size of the company it is difficult to hire specialist personnel for the sole purpose of executing these functions therefore as a compromise the company can employ this system to achieve some basic planning and monitoring.

## Business Objectives and Success Criteria

## Business Objectives:

BO-1: Reduce operational costs to within 80% of total revenue by elimination of non essential business travel, communication costs and the need for specialist roles within a year
Scale: Operational costs
Meter: Examination of financial reports
BO-2: Increase average effective work time by 20 minutes per employee per day within 3 months following initial release.
Scale: Ratio of labor to work load
Scale: Job rotors

## Success Criteria:

SC-1: Have up to 90% of all business transacted through the online management system within 12 months
SC-2: Realize a 90% customer satisfaction rating within 3 months
SC-3: Achieve a 90% customer retention rate within 3 months

## Business Risks

RI-1: Too few customers may employ the system
RI-2: It is likely that too many employees might appeal their ratings resulting into increased managerial overhead
RI-3: Too few employees may use the system meaning the employee ratings might have no true impact on employee behavior
RI-4: Some suppliers might have their own online ordering systems which might result in interoperability issues

# Appendix 2 ï Vision and Features

## Vision of the Solution
## Vision Statement

For a small business looking for an intuitive way to reduce costs, improve coordination, motivation, planning and monitoring this system provides a low cost internet based solution. It can be accessed via the companyôs website by any computer that is connected to the internet. Certain modules like the financial and accounting modules do not strictly require internet connectivity though the system as a whole requires internet connectivity to operate.

## Major Features

FE-1: Order products from suppliers
FE-2: Customers order services
FE-3: Create, view, modify, and delete user accounts
FE-4: Population of attendance sheets by supervisors and managers
FE-5: Checking for work rotors by employees
FE-6: Automatic scheduling of jobs and allocation of labor
FE-7: Definition of promotions by suppliers
FE-8: Rating of services by customers
FE-9: Rating of employees by supervisors
FE-10: Automatic ranking of employees based on productivity
FE-11: Appealing of a rating by an employee
FE-12: Reporting functions
FE-13: Printing of reports
FE-14: Automatic job scheduling
Assumptions and Dependencies
AS-1: Users will have access to internet-enabled computers and printers

Scope of Initial and Subsequent Releases

| Feature | Release 1 | Release 2 | Release 3 |
|---|---|---|---|
| FE-1 | Fully implemented | | Mobile version |
| FE-2 | Fully implemented | Mobile version | |
| FE-3 | Fully implemented | | |
| FE-4 | Fully implemented | | |
| FE-5 | Fully implemented | | Mobile version |
| FE-6 | Fully implemented | | |
| FE-7 | Fully implemented | | |
| FE-8 | Fully implemented | Review ratings algorithm | Review ratings algorithm |
| FE-9 | Fully implemented | | |
| FE-10 | Fully implemented | Review rankings algorithm | Review rankings algorithm |
| FE-11 | Fully implemented | | |
| FE-12 | Fully implemented | Increments: | Mobile version |
| FE-13 | Fully implemented | | |
| FE-14 | Fully implemented | Review scheduling algorithm | Review scheduling algorithm |

## Limitations and Exclusions

LI-1: Financial reports are limited to business transacted through the system

# Appendix 3ï Business Context

## Project Priorities

| Dimension | Driver | Constraint | Degree of Freedom |
|---|---|---|---|
| Schedule | | | |
| Features | | All features scheduled for release 1.0 must be fully operational | |
| Quality | | Must pass up to 90% of user acceptance tests; all security tests must pass; compliance with company security standards must be demonstrated for all secure transactions, all transactions must adhere to applicable legislation, | |
| Staff | projected team size is half-time project manager, 3 programmers, and half-time tester; additional half-time programmer will be available if necessary | | |
| Cost | | | budget overrun up to 10% acceptable, requires clientôs review |

The next step is the formulation of business rules. The table below outlines the business rules applicable to this business.

# Appendix4 ï Business Rules

| Rule ID | Rule Description | Type of Rule | Static or Dynamic | Source |
|---|---|---|---|---|
| BR1 | All system users are required to have an account | Fact | static | Company security policy |
| BR2 | Users can not access system whilst on suspension or after leaving the company | Fact | static | Company security policy |
| BR3 | Product re-order level is 50% of store capacity | Fact | dynamic | Company operational procedures |
| BR4 | Managers and supervisors may only access the attendance sheets of shifts in which they are designated as ñon dutyò | Constraint | static | Company security policy |
| BR5 | The cost of a one off service is calculated as the cost of labor plus the cost of cleaning products plus the applicable sales tax plus a given profit margin | Computation | dynamic | Company pricing strategy, tax code |
| BR6 | Managers and supervisors may only access the inventory sheets of shifts in which they are designated as ñon dutyò | Constraint | static | Company security policy |
| BR7 | In case of scheduling conflicts contracted customers take priority | Constraint | static | Company operational procedures |
| BR8 | Network transmissions that involve financial information or personally identifiable information require 128-bit encryption. | Constraint | Static | Company security policy |
| BR9 | Users are required to change their passwords every three weeks | Fact | Static | Company security policy |
| BR10 | Cash flow statements are generated for each running month | Fact | Static | Company financial policy |
| BR11 | Attendance and hourly productivity reports are generated weekly for each employee | Fact | Static | Company Human Resource policy |
| BR12 | Inventory summaries are required on a daily basis | Fact | Static | Company operational procedures |
| BR13 | Unscheduled, adhoc financial and inventory reports are occasionally required for real time business evaluation | Fact | Static | Company operational procedures |
| BR14 | Order price is calculated as the sum of item price times the quantity of that item, plus applicable sales tax, plus a delivery charge | Computation | Dynamic | Supplier pricing strategy; state tax code |
| BR15 | All cleaning products in a single order must be delivered to the same location. | Constraint | Static | Supplierôs policy |
| BR16 | All cleaning products in a single order must be paid for using the same payment method. | Constraint | Static | Supplierôs policy |
| BR17 | A delivery charge that is calculated according to the supplierôs contract terms is placed on deliveries | Computation | Dynamic | Supply Contract terms |
| BR18 | Only the director or administrator may add, delete, modify or suspend a user account | Constraint | static | Company security policy |
| BR19 | Should any changes be made to a user account by the administrator, they should be authorized or requested by the director | Constraint | static | Company security policy |
| BR20 | Customers can only rate those jobs that they requested and were completed | Constraint | static | |
| BR21 | | | | |

# Appendix 5 ï Full System Requirements Specification (SRS)

## Introduction

## Purpose
This section describes the software functional and nonfunctional requirements for release 1.0 of the Online Management System for Keymaid Cleaning Agency. The contents of this section are intended for members of the project team that will implement and test the system. All requirements included here are intended for release 1.0 and are high priority.

.

## User Classes and Characteristics

**Customer** The customerõs request services through the system via the internet. They are required to have an account in order to request and pay for services. They can also optionally rate the quality of the service for a given completed job and view the history of all jobs that have been completed.

**Administrator** Manages the database and is the person who is contacted in case there is a problem that users canõt independently resolve

**Supplier** The suppliers are those external businesses that supply the company with cleaning products. Most of the relationships with suppliers are automatically managed but occasionally it is necessary for the director to make one off product requests. These requests are executed through the system but depending on the current usage of products the system automatically sends requests to the suppliers via emails to replenish the companyõs stock

**Director** A director is the owner of the small business or someone operating in his place. The director is the highest position in the business and almost possesses administrative privileges. The director has the right to fire, recruit, promote, demote and suspend employees and this is mirrored in the system by his ability to delete, add, modify and suspend accounts. In addition the director may access any report generated by the system and will be the only one with rights to make product orders for cleaning products and authorize payments. The system provides the director with a bird eyeõs view of the business

**Cleaning Staff** The Keymaid business at the moment employs five cleaning staff. The number is expected to increase threefold by the end of the first year and they are expected to be computer literate. They are charged with executing the customer requests and they will use the system to access rotors, individual rankings and appeal ratings

## Operating Environment

**OE-1:** The System shall operate with the following Web browsers: Microsoft Internet Explorer versions 5.0 and above, Safari browser, Mozilla firefox, Opera

**OE-2:** The System shall operate on a server running Windows Server 2000 and Apache Web Server.

**OE-3:** The System shall permit user access via the internet
Design and Implementation Constraints

**CO-1:** The systemĝs design, code, and maintenance documentation shall conform to the Keymaid Intranet Development Standard, Version 1.3 [2].

**CO-2:** The system shall use Mysql

**CO-3:** All HTML code shall conform to the HTML 4.0 standard.

**CO-4:** All scripts shall be written in php

## User Documentation

**UD-1:** The system provides an online hierarchical and cross-linked help system in HTML that describes and illustrates all system functions.

## Assumptions and Dependencies

**AS-1:** The business operates 24 hours a day, 7 days a week

**DE-1:** The operation of the system depends on the state of suppliersô systems and operations
System Features

## Use cases

The following use cases depict a possible requirements analysis for an internet based application for Keymaid Cleaning Agency:

| Primary Actor | Use Cases |
|---|---|
| Director | Order cleaning products |
| Director/ Manager | Run summary reports |
| Supplier | Modify Menu<br>Define promotions |
| Customers | Order one off service<br>Create account<br>Rate Service |
| Administrator/ Director | Delete or suspend Usersô accounts for Online Management System |
| Supervisor | Populate cleanersô attendance sheet<br>Populate cleaning productsô inventory sheets |
| Use Case ID: | 1 |
| Use Case Name: | Order cleaning products |
| Actors: | Director |
| Description: | The director accesses the Online Management System from the corporate intranet or from home, optionally views the inventory summaries, selects cleaning/ office products in need of replenishment, and places an order for a product to be delivered to the companyĝs site |
| Preconditions: | Director is logged into Online Management System. |
| Postconditions: | Product order is stored in Online Management System with a status of ñsentò.<br>Product inventory is updated to reflect items in this order.<br>Remaining store capacity is updated to reflect this delivery request. |
| Normal Flow: | 1.0 Order a Single Product<br>1. Director asks to view product menu for a given date<br>2. System displays menu of products by supplier<br>3. Director selects one or more cleaning/ office products from menu indicating the amount<br>4. Director indicates that product order is complete.<br>5. System displays ordered menu items, individual prices, and total price, including any taxes and delivery charge.<br>6. Director confirms product order or requests to modify product order (back to step 3).<br>7. System displays available delivery times for the delivery date.<br>8. Director selects a delivery time and specifies the delivery location.<br>9. Director specifies payment method.<br>10. System confirms acceptance of the order.<br>11. System sends Director an e-mail confirming order details, price, and delivery instructions.<br>12. System stores order in database, sends e-mail to notify Supplier |
| Alternative Flows: | 1.1 Order multiple products (branch after step 4)<br>Director asks to order another product.<br>Return to step 2. |
| Exceptions: | 1.0.E.1 Current time is after order cutoff time (at step 1)<br>1. System informs Director that itôs too late to place an order for today.<br>2a. Director cancels the product order.<br>2b. System terminates use case.<br>3a. Director requests to select another date.<br>3b. System restarts use case.<br><br>1.0.E.2 No delivery times left (at step 1)<br>1. System informs Director that no delivery times are available for the product date.<br>2a. Director cancels the product order.<br>2b. System terminates use case.<br>3. Director requests to pick the order up at the supplier (skip steps 7-8).<br><br>1.2.E.1 Canôt fulfill specified number of identical products (at step 1)<br>System informs Director of the maximum number of identical products it can supply.<br>Director changes number of identical products ordered or cancels product order. |
| Includes: | None |
| Priority: | High |
| Frequency of Use: | 1 user, average of one usage per week (products ordered weekly) |
| Business Rules: | 1, 8, 9, 14 |
| Special Requirements: | Director shall be able to cancel the product order at any time prior to confirming the order.<br>Director shall be able to view all products he ordered within the previous six months and repeat one of those products as the new order, provided that all products are available on the menu for the requested delivery date. (Priority = medium) |
| Notes and Issues: | The default date is the current date if the Director is using the system before todayôs order cutoff time. Otherwise, the default date is the next day that the supplier is open. |
| Use Case ID: | 2 |
| Use Case Name: | Run summary reports |
| Actors: | Director |
| Description: | The director accesses the Online Management System from the corporate intranet or from home selects his/her profile and optionally to view reports |
| Preconditions: | Director is logged into Online Management System. |
| Postconditions: | The generated report is dated with the current date and stored for future reference |
| Normal Flow: | 1. Run summary reports<br>2. Director asks to view reports<br>3. System displays menu of report categories<br>4. Director selects a single report category<br>5. System displays available reports within the category<br>6. Director selects a single report<br>7. System runs the report<br>8. System asks if the director wishes to run another report<br>9. Director says ñyesò<br>10. System returns to step 2 |
| Alternative Flows: | Director chooses not run another report (branch at step 8)<br>System terminates use case |
| Includes: | None |
| Priority: | High |
| Frequency of Use: | Several times a day |
| Business Rules: | 1, 8, 9, 10, 11, 12, 13 |
| Special Requirements: | Director shall be able to print or email reports<br>Director shall be able to view all reports he has run within the previous six months |
| Use Case ID: | 3 |
| Use Case Name: | Modify Menu |
| Actors: | Supplier |
| Description: | The Supplier may modify the menu of available products and prices for a specified date to reflect changes in availability or prices |
| Preconditions: | Menus already exist in the system. |
| Postconditions: | Modified menu has been saved. |
| Normal Flow: | 11.0 Edit Existing Menu<br>1. Supplier requests to view the menu for a specific date.<br>2. System displays the menu.<br>3. Supplier modifies the menu to add new products, remove or change products, create or change an ongoing promotion, or change prices.<br>4. Supplier requests to save the modified menu.<br>5. System saves modified menu. |
| Alternative Flows: | None |
| Exceptions: | 11.0.E.1 No menu exists for specified date (at step 1)<br>1. System informs Supplier that no menu exists for the specified date.<br>2. System asks Supplier if he would like to create a menu for the specified date.<br>3a. Supplier says yes.<br>3b. System invokes Create Menu use case.<br>4a. Supplier says no.<br>4b. System terminates use case.<br><br>11.0.E.2 Date specified is in the past (at step 1)<br>System informs Supplier that the menu for the requested date cannot be modified. |
| | System terminates use case. |
| Includes: | Create Menu |
| Priority: | High |
| Frequency of Use: | Approximately 20 times per week by one user |
| Business Rules: | BR-24 |
| Special Requirements: | The Supplier may cancel out of the menu modification function at any time. If the menu has been changed, the system shall request confirmation of the cancellation. |
| Assumptions: | A menu will be created for every official Keymaid business day, including weekends and holidays in which employees are scheduled to be on site. |
| Use Case ID: | 4 |
| Use Case Name: | Define Promotion |
| Actors: | Supplier |
| Description: | The Supplier may define a promotion by populating adding products to the promotions menu |
| Preconditions: | Supplier is logged into the system<br>Product is listed in the generic menu |
| Postconditions: | New promotion is added to the promotions menu<br>Product is listed as being on promotion in the generic product menu<br>Product is listed with the old price and promotion price |
| Normal Flow: | 1.0 Define promotion<br>1. Supplier optionally requests to create new promotion or chooses existing promotion<br>2. System prompts supplier for promotion name and start and end dates<br>3. Supplier enters the promotion name along with start and end dates<br>4. System generates checklist of items by the given supplier<br>5. Supplier checks off items to add to the given promotion<br>6. Supplier confirms product choices by choosing ñcontinueò<br>7. System generates list of items in given promotion with current prices and prompts supplier for new prices<br>8. Supplier confirms entries by choosing ñcontinueò<br>9. System generates preview menu for new promotion and prompts supplier for confirmation<br>10. Supplier confirms entries by choosing ñyesò option<br>11. System saves new menu<br>12. System asks supplier if he wishes to create another promotion<br>13. If supplier chooses ñnoò system terminates use case |
| Alternative Flows: | 1.0. Supplier chooses to edit existing promotion (branch at step 1)<br>System generates checklist of promotions<br>Supplier checks off a single promotion to edit<br>Supplier confirms choice by choosing ñcontinueò<br>Repeat steps 7 to 11 above<br>System terminates use case |
| Exceptions: | 1.0.E.1 No promotion exists (at step 1, alternative flow)<br>1. System informs Supplier that no promotion exists<br>2. System asks Supplier if he would like to create a new promotion<br>3a. Supplier says yes.<br>3b. System invokes Create promotion use case.<br>4a. Supplier says no.<br>4b. System terminates use case.<br><br>11.0.E.2 Date specified is in the past (at step 2)<br>System informs Supplier that the menu for the requested date cannot be modified.<br>System terminates use case. |
| Includes: | Create promotion |
| Priority: | High |
| Frequency of Use: | Approximately 10 times per year by a given supplier |
| Business Rules: | 1, 8, 9 |
| Special Requirements: | The Supplier may cancel out of the promotion modification function at any time. If the promotion has been changed, the system shall request confirmation of the cancellation. |
| Use Case ID: | 5 |
| Use Case Name: | Order one-off Service |
| Actors: | Customer |
| Description: | The customer accesses the Online Management System externally via the companyôs website optionally views service offers, a given service, and places an order for a job to be delivered to the companyôs site |
| Preconditions: | Customer is logged into Online Management System. |
| Postconditions: | Requested services are stored in the customerôs profile area on the Online Management System with a status of ñpendingò.<br>Job sheets are updated to reflect the requested services.<br>The companyôs remaining capacity to satisfy additional requests is updated to reflect the requested service |
| Normal Flow: | 1.0 Order one-off Service<br>1. Customer asks to view the services menu for a given date<br>2. System displays menu of available services by the agency<br>3. Customer selects one or more services from menu and confirms by choosing ñcontinueò<br>4. If customer is not logged in the system invokes ñauthenticate userò case<br>5. System prompts customer for date, time and location for the requested service<br>6. Customer enters date, time and location<br>7. System verifies whether the company has enough capacity to fulfill the customerôs request<br>8. Customer indicates that a service order is complete.<br>9. System displays ordered menu items, individual prices, and total price, including any taxes and delivery charge.<br>10. Customer confirms service order or requests to modify service order (back to step 3).<br>11. Customer specifies payment method.<br>12. System confirms acceptance of the order.<br>13. System sends Customer an e-mail confirming order details, price, and delivery instructions.<br>14. System stores order in database, sends e-mail to notify Supplier |
| Alternative Flows: | 1.1 Order multiple jobs (branch after step 4)<br>Customer asks to order another job.<br>Return to step 2.<br><br>1.2 Order multiple identical jobs (after step 3)<br>Customer requests a specified number of identical jobs.<br>Return to step 4. |
| Exceptions: | 1.0.E.2 Company does not have enough capacity to fulfill the customer request (at step 6)<br>1. System informs Customer that it doesnôt have enough capacity to fulfill the job<br>2a. Customer cancels the job order.<br>2b. System terminates use case.<br>3. Customer requests to pick the order up at the supplier (skip steps 7-8). |
| Includes: | None |
| Priority: | High |
| Frequency of Use: | 20 orders per week, more users with time |
| Business Rules: | 1, 5, 8, 9, |
| Special Requirements: | Customer shall be able to cancel the job order at any time prior to confirming the order.<br>Customer shall be able to view all jobs he ordered within the previous six months and repeat one of those jobs as the new order, provided that all jobs are available on the menu for the requested delivery date. (Priority = medium) |
| Use Case ID: | 6 |
| Use Case Name: | Create account |
| Actors: | Customer |
| Description: | A customer may access the system via the company website from anywhere |
| Postconditions: | A new profile is created for the given customer detailing the customersô transactions with the company |
| Normal Flow: | 1.0 Create Account<br>1. Customer requests to create account<br>2. Customer prompted for email<br>3. Customer asked to confirm email<br>4. Customer optionally enters payment details<br>5. Customer confirms entries by selecting ñcontinueò<br>6. Customer presented with terms of use<br>7. Customer asked if they agree to terms of use<br>8. Customer accepts by selecting ñI Agreeò<br>9. An activation email containing activation password is sent to the customerõs email address<br>10. Customer redirected to account activation page via link in activation email<br>11. Customer prompted for email address and activation password<br>12. System invokes ñauthenticate userò case<br>13. Customer redirected to profile page |
| Alternative Flows: | None |
| Exceptions: | 1.0.E.1 Emails entered by user do not correspond (at step 3)<br>System informs customer that the emails do not match and prompts the user to try again<br>Continue to step 4<br>2.0.E.2 User Authentication Fails (at step 9)<br>System informs customer that login attempt has failed<br>Customer asked to enter details again and is reminded that password is case sensitive<br>If details are correct on second or third attempt the customer account is saved otherwise the use case is terminated and customer advised to repeat the attempt |
| Includes: | Authenticate user |
| Priority: | High |
| Frequency of Use: | Once for each customer |
| Business Rules: | BR-24 |
| Assumptions: | Any customer that does a transaction with the company requires an account. |
| Use Case ID: | 7 |
| Use Case Name: | Rate Service |
| Actors: | Customer |
| Description: | The Customer may rate a given job to indicate the extent to which he is satisfied with the companyõs service via a confidential form accessible through his profile page |
| Preconditions: | Customer is logged into the system |
| Normal Flow: | 1.0 Rate Service<br>1. Customer optionally selects to rate a given job<br>2. System displays checklist of completed unrated jobs in the customerõs account<br>3. Customer checks off job to be rated<br>4. System generates feedback form to be completed by the customer |
| | Customer indicates completion of form by selecting ñsubmitò<br>System terminates use case |
| Alternative Flows: | None |
| Exceptions: | 1.0.E.1 No completed, unrated jobs exist for the customer account (at step 2)<br>System informs the customer that there are no completed jobs to be rated<br>System terminates use case |
| Includes: | None |
| Priority: | High |
| Frequency of Use: | Approximately 10 times per year by a given customer |
| Business Rules: | BR-24 |
| Special Requirements: | The Customer may cancel out of the job rating function at any time.<br>If the feedback form has been changed, the system requests a confirmation before cancellation<br>The system allows the customer to save the form for further completion<br>The feedback form must be printable |
| Use Case ID: | 8 |
| Use Case Name: | Delete or suspend Usersô accounts for Online Management System |
| Actors: | Administrator/ Director, employee database |
| Description: | In case any member of staff is lost or on suspension their accounts are deleted from the system |
| Preconditions: | User is logged into the Online Management System. |
| Normal Flow: | 6.0 Delete or suspend usersô accounts from online management system<br>1. User requests to delete or suspend a userôs account<br>System invokes Authenticate Userôs Identity use case.<br>User is prompted for the employees name<br>System invokes search user use case<br>User chooses a single employee<br>User is transferred to the employees profile<br>User chooses the delete option<br>User requested for confirmation by system<br>User gives confirmation<br>System displays confirmation message to the user |
| Alternative Flows: | 1.1. User requests to suspend a userôs account (branch at step 7)<br>User chooses the suspend option<br>User is prompted for start and end date<br>User is prompted for end date<br>User is prompted for confirmation<br>Return to step 9 |
| Exceptions: | 6.0.E.1 User identity authentication fails (at step 2)<br>1. System gives user two more opportunities for correct identity authentication.<br>2a. If authentication is successful, User proceeds with use case.<br>2b. If authentication fails after three tries, System notifies user, logs invalid authentication attempt, and terminates use case. |
| Includes: | Authenticate Userôs Identity |
| Priority: | High |
| Frequency of Use: | Thrice a year |
| Business Rules: | 1, 2, 8, 9, |
| Special Requirements: | User authentication is performed per corporate standards for medium-security applications. |
| Assumptions: | None |
| Use Case ID: | 9 |
| Use Case Name: | Populate Cleaning products inventory sheets |
| Actors: | Supervisor |
| Description: | The supervisors are to populate inventory sheets for cleaning products so that the system can keep track of inventory |
| Preconditions: | Supervisor is logged into the system |
| Postconditions: | Product inventory is adjusted to reflect the actual inventory as entered by the supervisor |
| Normal Flow: | 1.0 Populate Cleaning products inventory<br>Supervisor selects inventory sheets<br>System generates a checklist of cleaning products<br>Supervisor confirms choice by selecting continue<br>System generates list of chosen items and prompts supervisor for amounts issued and amounts returned<br>System automatically works out amounts used and the standing inventory<br>System displays inventory summary indicating amounts issued, amounts used and standing inventory and prompts supervisor for confirmation<br>Supervisor gives confirmation by choosing ñcontinueò<br>System saves data<br>System terminates use case |
| Alternative Flows: | None |
| Includes: | None |
| Priority: | High |
| Frequency of Use: | Once for each shift that is an attendance sheet is generated for each job |
| Business Rules: | 1, 6, 8, 9 |
| Special Requirements: | Supervisor must be able to recall previous inventory sheets for up to 6 months<br>Inventory sheets should be printable<br>Inventory sheets are modifiable |
| Use Case ID: | 10 |
| Use Case Name: | Share |
| Actors: | customer |
| Description: | The customer can share a product description or rating by posting it to an external website via the defined APIS |
| Preconditions: | User is logged in to the system |
| Postconditions: | None |
| Normal Flow: | 1.0 Share<br>1. User selects product/ service<br>2. User expands details tab<br>3. User selects ñshareò<br>4. User selects target website e.g. Facebook.com, Delicious |
| Alternative Flows: | None |
| Includes: | None |
| Priority: | High |
| Frequency of Use: | Once for each shift that is an attendance sheet is generated for each job |
| Business Rules: | 1, 6, 8, 9 |
| Special Requirements: | Supervisor must be able to recall previous inventory sheets for up to 6 months<br>Inventory sheets should be printable<br>Inventory sheets are modifiable |

## Functional Requirements

## Order Cleaning Products

### 3.1.1 Description and Priority

The system automatically sends requests for replenishment of stock of cleaning products depending on the inventory rules or in occasional circumstances the director may place an order for stock, high priority

### 3.1.2 Stimulus/Response Sequences

Stimulus: Director requests to place an order for cleaning products or the systemôs inventory diminishes to a threshold value (as per BR3) for systemôs re-order function
Response: System queries director for details of product(s), payment, and delivery instructions.
Stimulus: Director requests to change a product order.
Response: If status is ñAccepted,ò system allows user to edit a previous product order.
Stimulus: Director requests to cancel a product order.
Response: If status is ñAccepted, òsystem cancels a product order.

3.1.3 Functional Requirements

| |
|---|
| Order.Place: The system shall let a Director who is logged into the System place an order for one or more products.<br>Order.User.Authenticate: The system shall confirm that the said director has a valid account<br>Order.Place.Date: The system shall prompt the director for the product delivery date<br>Order.Place.Date.Cutoff: If the product date is the current date and the current time is after the order cutoff time, the system shall inform the director that itôs too late to place an order for today. The Director may either change the product date or cancel the order. |
| Order.Deliver.Select: The Director shall specify whether the order is to be picked up or delivered.<br>Order.Deliver.Location: If the order is to be delivered and there are still available delivery times for the product date, the director shall provide a valid delivery location.<br>Order.Deliver.Notimes: The system shall notify the director if there are no available delivery times for the product date. The director shall either cancel the order or indicate that the company will pick up the products from the supplier<br>Order.Deliver.Times: The system shall display the remaining available delivery times for the product date. The system shall allow the director to request one of the delivery times shown, to change the order to be picked up at the supplier, or to cancel the order. |
| Order.Menu.Date: The system shall display a menu for the specified date.<br>Order.Menu.Available: The menu for the current date shall display only those cleaning products for which at least one item is available in the supplierôs product inventory |
| Order.Units.Product: The system shall allow the Director to indicate the number of units of each menu item that he wishes to order.<br>Order.Units.Multiple: The system shall permit the user to order multiple identical units, up to the fewest available units of any menu item in the order.<br>Order.Units.TooMany: If the Director orders more units of a menu item than are presently in the supplierôs inventory, the system shall inform the director of the maximum number of units of that cleaning product that he can order.<br>Order.Units.Change: If the available inventory cannot fulfill the number of units ordered, the Director may change the number of units ordered, change the number of identical products being ordered, or cancel the cleaning product order. |
| Order.Confirm.Display: When the director indicates that he does not wish to order any more cleaning products, the system shall display the cleaning products ordered, the individual item prices, and the payment amount, calculated per BR-14, BR15, BR16, BR17<br>Order.Confirm.Prompt: The system shall prompt the Director to confirm the product order.<br>Order.Confirm.Not: If the Director does not confirm the product order, the Director may either edit or cancel the order.<br>Order.Confirm.More: The system shall let the Director order additional products for the same or for different date. BR-15 and BR-16 pertain to multiple cleaning products in a single order. |
| Order.Pay.Method: When the director indicates that he is done placing orders, the system shall ask the user to select a payment method.<br>Order.Pay.Deliver: See BR-17.<br>Order.Pay.Details: The system shall display the cleaning products ordered, payment amount, payment method, and delivery instructions.<br>Order.Pay.Confirm: The director shall either confirm the order, request to edit the order, or request to cancel the order.<br>Order.Pay.Confirm.OK: If the payment request is accepted, the system shall display a message confirming acceptance of the order with the receipt number<br>Order.Pay.Confirm.NG: If the payment request is rejected, the system shall display a message with the reason for the rejection. The director shall either cancel the order, or change the payment method to cash and request to pick up the order from the supplier |
| Order.Done: When the Director has confirmed the order, the system shall do the following as a single transaction:<br>Order.Done.Store Assign the next available cleaning product order number to the cleaning product and store the cleaning product order with an initial status of ñAccepted.ò<br>Order.Done.Inventory: Send a message to the inventory system with the type and number of units in the given order<br>Order.Done.Menu: Update the menu for the current orderôs order date to reflect any items that are now out of stock in the suppliersô inventory.<br>Order.Done.Times: Update the remaining available delivery times for the date of this order.<br>Order.Done.Director: Send an e-mail message to the Director with the cleaning product order and cleaning product payment information.<br>Order.Done.Supplier: Send an e-mail message to the Supplier with the cleaning product order information.<br>Order.Done.Failure: If any step of Order.Done fails, the system shall roll back the transaction and notify the user that the order was unsuccessful, along with the reason for failure. |
| Order.Previous.Period: The system shall permit the Director to view any cleaning products he has ordered within the previous six months. [Priority = Medium]<br>Order.Previous.Reorder: The Director may reorder any cleaning product he had ordered within the previous six months, provided that all food items in that order are available on the menu for the cleaning product date. [Priority = Medium] |

## Create a user accounts

3.2.1 Description and Priority

All system users are required to have an account before using the system. Company users are given accounts by the administrator or director and their accounts may be added, deleted, modified, suspended or deleted to reflect changes in staff like hiring, suspension, lateral moves, termination and promotion or demotion. Customers can create their own accounts via the internet. High priority

3.2.2 Stimulus/Response Sequences

Stimulus: Director or administrator or customer requests to create a user account as per BR 1, BR18, BR19
Response: System queries director for details of employee identification, department and position
Stimulus: Director requests to delete an employee account
Response: System queries director for details of employee identification, department and position
Stimulus: Director requests to suspend an employee account
Response: System queries director for details of employee identification, department and position

3.2.3 Functional Requirements

| |
|---|
| Account.create The system will let the director or administrator or customer who is logged in to the system<br>create a new employee account |
| Account.User.authenticate:<br>The system will establish whether the current user has rights to create an account |
| Account.create.userName:<br>The system will prompt the user to enter the email address of the account Holder for customers and for company employees it will automatically generate a user name by appending the first letter of the employee's first name to the employee's second name |
| Account.create.Password<br>The system will prompt the user to enter the password for the account if the user is a customer or automatically generate a provisional password for the new employee which corresponds to a unique employee recruitment number |
| Account.create.Department<br>For employee accounts the system shall request the employee's department |
| Account.create.Role: For employee accounts the system shall request the employee's position Within the department |
| Account.Done: When the user has confirmed creation of the account the system will do the Following as a single transaction |
| Account.Done.Store: The new user account is added to the appropriate database for employee Accounts they are placed in a temporary database awaiting activation by the Given employees |
| Account.Done.IdlePane:<br>For employee accounts the new account is moved to the idle accounts pane awaiting activation by the employee |
| Account.Done.Confirm:<br>The system provides the user with a message confirmation that the account Has been created |
| Account.Done.Failure:<br>If any step of Account.Done fails, the system shall roll back the transaction and notify the user that the order was unsuccessful,along with the reason for failure |

## Modify a user account

3.3.1 Description and Priority

The director or administrator as per BR1, BR18 and BR19 may modify a user account to reflect changes in staff such as promotion, demotion and lateral moves

3.3.2 Stimulus/Response Sequences

Stimulus: Director requests to modify an employee's account.

Response: System queries director for details of employee identification, department and position

Functional Requirements

| |
|---|
| Account.modify The system will let the director or administrator who is logged in to the system modify an employee account |
| Account.User.authenticate:<br>The system will establish whether the current user has rights to modify or delete accounts |
| Account.modify.employeeID:<br>If the user logged on is the director or administrator, the system shall request for the name of the employee whose account is to be deleted or modified |
| Account.modify.employeeSearch<br>The system shall search the employee database for the employee whose account is to be modified Using the employee id entered |
| Account.modify.role<br>The system shall request the user to choose from a menu of new employee roles for the employee |
| Account.modify.Done:<br>When the user has confirmed creation of the account the system will do the following as a single transaction |
| Account.modify.Done.Save: The employee details are copied to the data structure that corresponds to the new employee category |
| Account.modify.Done.Confirm: The system provides the user with a message confirmation that the account has been modified displaying the new employee profile |
| Account.modify.Done.Failure: If any step of Account.modify.Done fails, the system shall roll back the transaction and notify the user that the order was unsuccessful, along with the reason for failure |

## Delete/Suspend user account

3.4.1 Description and Priority

The director or administrator as per BR1, BR18 and BR19 may delete a user account to reflect changes in staff such as termination, resignations and in addition they may have to suspend accounts in case of staff suspensions. Customers also have the option to terminate their accounts

3.4.2 Stimulus/Response Sequences

Stimulus: Director requests to modify an employeeês account.

Response: System queries director for details of employee identification, department and position

Functional Requirements

| |
|---|
| Account.delete/ Account.suspend The system will let the director or administrator or customer who is logged in to the system delete an account and in addition administrator or director may suspend an account |
| Account.User.authenticate: The system will establish whether the current user has rights to modify or delete accounts |
| Account.delete.employeeID/ Account.suspend.employeeID: If the user logged on is the director or administrator, the system shall request for the name of the employee whose account is to be deleted or modified |
| Account.suspend.startDate: If the account is to be suspended, then the system will request the start date for the suspension |
| Account.suspend.startDate: If the account is to be suspended, then the system will request the end date for the suspension |
| Account.delete.employeeSearch/ Account.suspend.employeeID: The system shall search the employee database for the employee whose account is to be modified Using the employee id entered |
| Account.delete.Done/ Account.suspend.Done: When the user has confirmed deletion or suspension of the account the system will do the following as a single transaction |
| Account.delete.Done.Database: If it is an account to be deleted the system deletes the employee details from the employee database. |
| Account.suspend.Done.Database: If it is an account to be suspended the system invokes trigger events to revoke database permissions for the given user for the duration of the suspension |
| Account.delete.Done.Confirm/ Account.suspend.Done.Confirm The system provides the user with a message confirmation that the account has been modified displaying the new employee profile |
| Account.delete.Done.Failure/ Account.suspend.Done.Failure: If any step of Account.delete.Done/ Account.suspend.Done fails, the system shall roll back the transaction and notify the user that the order was unsuccessful, along with the reason for failure |

## Order Services

3.5.1 Description and Priority

The customer who is logged in via the internet may place an order for a service(s).

3.5.2 Stimulus/Response Sequences

Stimulus: Customer requests to place an order for a service(s)
Response: System queries director for details of product(s), payment, and delivery instructions.
Stimulus: Customer requests to change an order.
Response: If status is ñAccepted,ò system allows user to edit a previous order.
Stimulus: Customer requests to cancel an order.
Response: If status is ñAcceptedò, system cancels the order

3.5.3 Functional Requirements

Order.Place: The system shall let a Customer who is logged into the System place an order for one or more services
Order.User.Authenticate: The system shall confirm that the said customer has a valid account
Order.Place.Date: The system shall prompt the customer for the date he would like to be provided with the requested service

Order.Execute.Location: The customer shall provide a valid location for the service
Order.Execute.Times: The system shall display alternative times in which the company may be able to execute the given request as per available labor. The system shall allow the customer to request one of the alternatives shown or to cancel the order.

Order.Menu.Date: The system shall display a menu of executable services for the given date
Order.Menu.Available: The menu for the current date shall display only those cleaning services available for the given date as is dictated by staff and inventory conditions

Order.Premises.size: The customer will describe the size of the premises in terms of square feet, rooms and number of floors
Order.Premises.Description: The customer shall provide additional description of the premises in terms of the type of floor/walls, whether it is toilet, hall, indoors or outdoors, kitchen, windows or regular cleaning
Order.Premises.noCapacity: The system shall notify the customer if the company lacks the resources to execute the request at the desired place and location
Order.Premises.scaleDown: If the available staff and inventory conditions do not permit execution of the requested service the system will suggest a scale downed version of the order which suits the given conditions

Order.Confirm.Display: When the customer indicates that he does not wish to order any more cleaning services, the system shall display the cleaning services ordered, the individual item prices, and the payment amount
Order.Confirm.Prompt: The system shall prompt the Customer to confirm the service order.
Order.Confirm.Not: If the Customer does not confirm the service order, the Customer may either edit or cancel the order.
Order.Confirm.More: The system shall let the Customer order additional services for the same or for different date.

Order.Pay.Method: When the customer indicates that he is done placing orders, the system shall ask the user to select a payment method.
Order.Pay.Details: The system shall display the cleaning services ordered, payment amount, payment method, and delivery instructions.
Order.Pay.Confirm: The customer shall either confirm the order, request to edit the order, or request to cancel the order.
Order.Pay.Confirm.OK: If the payment request is accepted, the system shall display a message confirming acceptance of the order with the receipt number
Order.Pay.Confirm.NG: If the payment request is rejected, the system shall display a message with the reason for the rejection. The customer shall either cancel the order, or change the payment method to cash and request to pick up the order from the supplier

| |
|---|
| Order.Done: When the Customer has confirmed the order, the system shall do the following as a single transaction:<br>Order.Done.Store Assign the next available service order number to the order and store the it with an initial status of ñpendingò<br>Order.Done.Inventory: Send a message to the inventory system with the type and number of units set aside for completion of the given order<br>Order.Done.Menu: Update the menu for the available services for the order date to reflect the remaining available capacity for the company to execute orders<br>Order.Done.Customer: Send an e-mail message to the Customer with the service order and payment information.<br>Order.Done.Failure: If any step of Order.Done fails, the system shall roll back the transaction and notify the user that the order was unsuccessful, along with the reason for failure. |
| Order.Previous.Period: The system shall permit the Customer to view any services he has ordered within the previous six months. [Priority = Medium]<br>Order.Previous.Reorder: The Customer may reorder any cleaning service he had ordered within the previous six months, provided that all food items in that order are available on the menu for the cleaning service date. [Priority = Medium] |

## Rate Services

3.6.1 Description and Priority

The customer who is logged in will be able to rate the companyôs services on a job per job basis as per BR20

3.6.2 Stimulus/Response Sequences

Stimulus: Customer requests to rate a completed job

Response: System prompts customer to chose from a menu of completed jobs

Functional Requirements

| | |
|---|---|
| Service.Rate | The system requests the customer who is logged in to optionally rate the service on a job per job basis |
| Service.Rate.Job: | The system will request the customer to chose a completed job to rate |
| Service.Rate.Scale: | The system will provide the customer with a scale with which to generate a subjective rating for the given job |
| Service.Rate.Done: | When the user confirms that he has finished rating the job the system will execute the following as a single transaction |
| Service.Rate.Done.Database: | The itemized as well as total score along with the job identifier are stored in a specific data structure within the jobs database |
| Service.Rate.Done.Confirm | The system provides the user with a message confirmation that the action has been executed |
| Service.Rate.Done.Failure: | If any step of Service.Rate.Done.Failurefails, the system shall roll back the transaction and notify the user that the order was unsuccessful, along with the reason for failure |

## Populate Attendance Sheets

3.7.1 Description and Priority

The manager or supervisor who is logged in to the system will be able to populate attendance sheets as per BR4

3.7.2 Stimulus/Response Sequences

Stimulus: Manager/supervisor requests to fill in attendance sheets

Response: System prompts supervisor/ manager to chose from a menu of completed shifts for which their status is ñinchargeò

3.7.3 Functional Requirements

| | |
|---|---|
| Attendance.Populate: | The supervisor/manager requests to fill in attendance forms |

| |
|---|
| Attendance.shift<br>The system will request the supervisor/manager to chose a shift for<br>Which he is designated as ñin chargeò |
| Attendance.Employee.ReportingTime:<br>For each employee the manager/supervisor will provide the reporting time |
| Attendance.Employee.ReportingTime:<br>For each employee the manager/supervisor will provide the finishing time |
| Attendance.Done<br>When the user gives the confirmation that he has finished populating the attendance sheet the system does the following as a single transaction |
| Attendance. Done.Database<br>The system will save the attendance information in the appropriate data structure in the employee database where it may be accessed by the payroll system module |
| Attendance. Done.Wages:<br>Using predefined rates the system will compute the wages for each employee |
| Attendance.Done.Failure:<br>If any step of Attendance.Done fails, the system shall roll back the transaction and notify the user that the order was unsuccessful, along with the reason for failure |

## Populate Inventory Sheets

3.7.1 Description and Priority

The manager or supervisor who is logged in to the system will be able to populate inventory sheets as per BR6

3.7.2 Stimulus/Response Sequences

Stimulus: Manager/supervisor requests to fill in inventory sheets on a shift per shift basis

Response: System prompts supervisor/ manager to chose from a menu of completed shifts for which their status is ñinchargeò

3.7.3 Functional Requirements

| |
|---|
| Inventory.Populate:<br>The supervisor/manager requests to fill in inventory forms |
| Inventory.shift<br>The system will request the supervisor/manager to chose a shift for<br>Which he is designated as ñin chargeò |
| Inventory.Item.AmountIssued:<br>For each inventory item the manager/supervisor will provide the amount issued |
| Inventory.Item.AmountReturned:<br>For each inventory item the manager/supervisor will provide the amount returned |
| Inventory.Usage.Calculate:<br>From the amounts issued and amounts returned the system will compute the actual usage of each item |
| Inventory.Done<br>When the user gives the confirmation that he has finished populating the attendance sheet the system does the following as a single transaction |
| Inventory.Done.Database<br>The system will save the inventory information in the appropriate data structure in the product database where it may be accessed by the inventory management module |
| Inventory.Done.Cost:<br>Using the unit purchase cost for each item the system will compute the cost of the used inventory and store the results in the appropriate data structure in the sales data base where it can be accessed by the financial module |
| Inventory.Done.Failure:<br>If any step of Inventory.Done fails, the system shall roll back the transaction and notify the user that the order was unsuccessful, along with the reason for failure |

## Nonfunctional Requirements

## External Interface Requirements (APIS)

-Compatibility with facebook.com API
-Compatibility with Digg API
-Compatibility with Delicious API
-Compatibility with Delicious folkid

## User Interfaces

UI-1: The system shall provide a help link from each displayed page to explain how to use that page.
UI-2: The Web pages shall permit navigation and food item selection using the keyboard alone, in addition to using mouse and keyboard combinations.
Hardware Interfaces
None identified
Software Interfaces
SI-1: Supplier Inventory Systems
SI-1.1: The System shall transmit the quantities of products ordered to suppliersô systems through a programmatic interface.
SI-1.2: The system shall check the Suppliersô Inventory System to determine whether a requested cleaning product
SI-1.3: When the Suppliersô Systems notify the System that a specific cleaning item is no longer available, the System shall remove that food item from the menu for the current date.

## Communications Interfaces

CI-1: The System shall send an e-mail message to the Director to confirm acceptance of an order, price, and delivery instructions.
CI-2: The System shall send an e-mail message to the Director to report any problems with the meal order or delivery after the order is accepted.
CI-3: The System shall send an e-mail message to the Supplier indicating the items required and their quantities, delivery details and total cost
CI-4: The System shall e-mail a receipt to the customer as proof of payment for given order

## Performance Requirements

PR-1: The system shall accommodate 100 users during the peak usage with an estimated average session duration of 10 minutes.
PE-2: All Web pages generated by the system shall be fully downloadable in no more than 30 seconds over a 40KBps modem connection.
PE-3: Responses to queries shall take no longer than 7 seconds to load onto the screen after the user submits the query.
PE-4: The system shall display confirmation messages to users within 4 seconds after the user submits information to the system.

## Safety Requirements

No safety requirements have been identified.

## Security Requirements

SE-1: All network transactions that involve financial information or personally identifiable information shall be encrypted per BR-8
SE-2: Users shall be required to log in to the System for all operations except viewing product menus
SE-3: Access to attendance sheets and inventory sheets will be subject to BR 4 and BR 6
SE-4: The system shall permit cleaning staff to view only their own ratings and time table information, not that of other employees

## Software Quality Attributes

Availability-1: The System shall be available to users on the corporate Intranet and to dial-in users 99.9% of the time between 5:00am and midnight local time and 95% of the time between midnight and 5:00am local time.
Robustness-1: If the connection between the user and the system is broken prior to an order being either confirmed or canceled, the System shall enable the user to recover an incomplete order.

# Appendix 6: Budget Plan

The following table may be employed as a guide for budgeting for the project should the client choose to implement the design

| **Item** | **Cost** | **Mandatory?** |
|---|---|---|
| Domain Name | £0 to 1000s | Yes |
| Web Building Software | £0 to unlimited | Yes |
| Labor | £400 to £1000s | Yes |
| Hosting | £0 to £1000s per year (paid hosting recommended) | No |
| Search Engine Submission | £15 to £75 per year | No |
| Additional Tools | unlimited | No |
| Advertising | unlimited | No |